\documentclass[
aps,prd,
preprintnumbers,
amsmath,
amssymb,
nofootinbib,
preprintnumbers
]{revtex4}
\usepackage{here}
\usepackage{footnote}
\usepackage{comment,braket}
\usepackage{epsf}
\usepackage{amsmath}
\usepackage{graphics}
\usepackage{amsfonts}
\usepackage{amssymb}
\usepackage{latexsym}
\usepackage{color}
\usepackage{natbib}
\usepackage{graphicx}
\usepackage{hyperref}
\usepackage{array}

\newcommand{\bh}{\text{(BH)}}
\newcommand{\co}{\text{(C)}}
\newcommand{\out}{\text{out}}
\newcommand{\inn}{\text{in}}

\usepackage[svgnames]{xcolor}
\definecolor{phthaloblue}{rgb}{0.0, 0.06, 0.54}
\hypersetup{
    colorlinks=true,
    linkcolor=blue,
    citecolor=blue,
    filecolor=blue,
    urlcolor=blue,
    }

\begin{document}
\title{On the ease of excitation of black hole ringing:\\Quantifying the importance of overtones by the excitation factors}
%\title{On the Ease of Excitation of Quasinormal Modes for Kerr Black Holes:\\Quantifying the importance of overtones by the excitation factor}
%Detailed formalism and methodology
%Sasaki-Nakamura
%wave-packet reflection and fitting
\author{Naritaka Oshita}
\email{naritaka.oshita@riken.jp}
\preprint{RIKEN-iTHEMS-Report-21}
\affiliation{
  $^1$RIKEN iTHEMS, Wako, Saitama, Japan, 351-0198
}

\begin{abstract}
The excitation factors of black hole quasinormal modes quantify the \textit{ease of excitation} of the quasinormal modes and are independent of the source of perturbation. We compute the excitation factors of Kerr black holes up to the 20th overtone and find that the 4th, 5th, and 6th overtones have the first three highest excitation factors for intermediate and high spin parameters. This provides an independent confirmation of the importance of overtones that has been confirmed by the fitting data analysis of numerical relativity waveforms beginning around the strain peak amplitude.
\end{abstract}

\maketitle

\section{Introduction}
A distorted single black hole relaxes to a stationary black hole, and in the mean time, it emits gravitational wave (GW) radiation which is a superposition of the quasi-normal (QN) modes of the black hole
\begin{equation}
h_+ + i h_{\times} = \frac{1}{r} \sum_{lmn} C_{lmn} S_{lmn} (\theta) e^{im \phi} \exp{\left[ -i\omega_{lmn} (t-r^{\ast}) \right]},
\end{equation}
where $h_+$ and $h_{\times}$ are strain amplitudes for the two polarizations, $\omega_{lmn}$ is the QN frequency, and $C_{lmn}$ is the excitation coefficient\footnote{For review of the QN modes of the Kerr black hole, see e.g. Refs. \cite{Berti:2009kk,Konoplya:2011qq}.}. The indices $l$ and $m$ are the angular and azimuthal numbers, respectively, defined for the spin-weighted spheroidal harmonic (SWSH) function, $S_{lm} (\omega, \theta)$ \cite{Teukolsky:1972my,Teukolsky:1973ha,Press:1973zz}. The amplitude of each QN mode is also determined by the SWSH factor $S_{lmn} (\theta) \equiv S_{lm} (\omega_{lmn}, \theta)$. QN modes are numbered by non-negative integers, $n$, in ascending order of the damping rates. The least damped mode ($n=0$) is called the fundamental mode and the others ($n\geq 1$) are referred as the overtones. The complex values of QN frequencies are the location of the poles of the black hole Green's function \cite{Leaver:1986gd,Sun:1988tz,Andersson:1995zk,Glampedakis:2001js,Glampedakis:2003dn,Nollert:1992ifk,Andersson:1996cm,Nollert:1998ys} in the complex frequency plane, and there is an infinite number of QN modes for each $l$-mode in the Kerr spacetime \cite{Leaver:1985ax}. By virtue of the no-hair theorem of black hole \cite{Israel:1967wq,Carter:1971zc,Hawking:1971vc}, the complex QN frequencies are universal and depend only on the black hole's mass, $M$, and angular momentum, $Ma$. On the other hand, the excitation coefficient, $C_{lmn}$, depends on the source of the perturbation, and therefore, it is challenging to predict which QN mode is most highly excited for general cases. However, it is possible to understand the {\it tendency} of which QN modes are easy to excite regardless of the source of ringdown \cite{Sun:1988tz,Leaver:1986gd,Andersson:1995zk,Glampedakis:2001js,Glampedakis:2003dn,Berti:2006wq,Zhang:2013ksa}. The excitation coefficient is represented by the product of the following two factors: the quasi-normal excitation factor (QNEF) $E_{lmn}$ and source factor $T_{lmn}$
\begin{equation}
C_{lmn} = E_{lmn} T_{lmn}.
\end{equation}
The two factors will be defined and explained later in more detail, and here let us briefly explain the role of each factor qualitatively. The QNEF, $E_{lmn}$, is independent of the source of ringdown, and is therefore a universal quantity that quantifies the {\it ease of excitation} of QN modes. The source factor is determined by the initial data of a distorted black hole, and is hence challenging to estimate, especially, when the non-linear distortion of the black hole is involved, e.g., binary black hole (BBH) mergers. Although the complete prediction of $C_{lmn}$ are still challenging, the detailed investigation of the QNEFs for up to higher overtones is useful to predict which QN modes are easy to excite. Also, one can estimate the source factor as $T_{lmn} = C_{lmn} / E_{lmn}$ once obtaining $E_{lmn}$ and extracting the values of $C_{lmn}$ from a ringdown waveform.

The detections of GW signals \cite{LIGOScientific:2016aoc,LIGOScientific:2016sjg,LIGOScientific:2016dsl,LIGOScientific:2017bnn,LIGOScientific:2017vox,LIGOScientific:2017ycc,LIGOScientific:2018mvr,LIGOScientific:2020ibl} emitted from BBH mergers by LIGO \cite{LIGOScientific:2014qfs} and Virgo \cite{VIRGO:2014yos} have stimulated an interest in the determination of QN frequencies from the observations. The accurate extraction of the QN frequencies of remnant black holes from detected ringdown signals is important to test the black hole no-hair theorem and general relativity. The linear perturbation theory works at late times in the ringdown phase where the fundamental mode, the least damped QN mode, dominates the waveform. On the other hand, other overtones are exponentially suppressed with higher damping rates. Therefore, it had been thought that higher overtones would be less important, at least, in the purpose of testing the no-hair theorem or general relativity. However, the overtones exponentially grow as going back in time towards the early phase of ringdown and can overwhelm the fundamental mode, provided that the perturbation theory is still relevant at the early ringdown. Indeed, it was found that \cite{Giesler:2019uxc} the GW150914-like numerical relativity waveform (SXS:BBH:0305) \cite{Mroue:2013xna} before the strain peak can be accurately modeled by a superposition of the fundamental QN mode and the first seven overtones. That indicates the spacetime can be described by a linearly perturbed Kerr black hole as early as the strain peak. This surprising result was followed by other extensive studies \cite{Ferguson:2019slp,Isi:2019aib,Bhagwat:2019dtm,Bhagwat:2019bwv,Varma:2020nbm,Cook:2020otn,JimenezForteza:2020cve,Capano:2020dix,Bustillo:2020buq,Dhani:2020nik,Mourier:2020mwa,Isi:2020tac,Finch:2021iip,Capano:2021etf,Isi:2021iql,Forteza:2021wfq,Dhani:2021vac,Ota:2021ypb,Sago:2021gbq}. Also, it was recently confirmed \cite{Ma:2021znq} that the hierarchy of the mode amplitudes and the significance of the 4th and 5th overtones are insensitive to the initial condition of {\it superkick} configuration of BBHs (those with equal mass and anti-parallel spins). In spite of the supporting evidence indicating the importance of overtones, it is still an open question why the overtones are highly excited at the early ringdown. Especially, most of the previous studies investigating the GW ringdown signals sourced by BBH mergers have reported that including the first seven overtones are enough to model the early and late ringdown phase\footnote{Ref. \cite{Forteza:2021wfq} found that including overtones up to $n=6$ is sufficient on average for numerical relativity waveforms in the SXS catalog, and Ref. \cite{Mourier:2020mwa} found that including tones up to $n= (l+2)$ is sufficient for the perturbation of a Schwarzschild black hole resulting from the head-on collision of two non-spinning black holes.} \cite{Giesler:2019uxc,Bhagwat:2019dtm,Dhani:2020nik,Finch:2021iip}, but the clear reason for this has not been understood from the theoretical side.

In this paper, we compute the QNEFs of $l=m=2$ up to the 20th overtone to quantify the ease of excitation of QN modes and provide novel evidence showing the importance of overtones. In the former part of Sec. \ref{sec:QNmode_QNEF}, we briefly review the QN modes and QNEFs and explain how we compute them up to the 20th overtone. Also, we slightly modify the original definition of QNEF \cite{Zhang:2013ksa} so that it quantifies the ease of excitation in strain that is independent of the choice of perturbation variables. In Sec. \ref{sec:QNEF_result}, we show that the 4th, 5th, and 6th overtones have the three highest QNEFs for $0.3 \leq j \lesssim 0.9$ ($j \equiv a/M$), which is well consistent with the truncation at $n=7$ in the fitting of QN modes to GW ringdown waveforms as have been applied in the literature \cite{Giesler:2019uxc,Bhagwat:2019dtm,Dhani:2020nik,Finch:2021iip}. We also find an interesting relation between the anomalous behaviour of the 5th QN frequency (see e.g. \cite{Onozawa:1996ux}) and the spin dependence of the 5th QNEF for $j \gtrsim 0.9$. We show that in the near-extremal situation, the value of the 5th QNEF is suppressed and the 5th QN mode becomes hard to excite. In Sec. \ref{sec:plungeGW}, we perform the numerical computation of a GW signal induced by a particle plunging into a black hole by solving the Sasaki-Nakamura equation \cite{Sasaki:1981sx} and extract the excitation coefficients $C_{22n}$ up to higher overtones. Then we check how much each overtone is excited from the extracted $C_{22n}$ and confirm that the 4th, 5th, and 6th overtones are highly excited for medium and high spin parameters. Also, we find that the 5th overtone is highly suppressed for the near-extremal case. This analysis is independent of the computation of the QNEFs, but nevertheless it is well consistent with the behaviour of the QNEFs computed in Sec. \ref{sec:QNEF_result}. In Sec. \ref{sec:decaytime}, we introduce a new physical quantity, $t_{lmn}$, that is useful to predict the time when the $n$-th overtone mode tends to be less dominant compared to the fundamental mode. The quantity $t_{lmn}$, referred as the {\it decay time}, is determined only by the QN frequency and QNEF, and so it is independent of the source of ringdown. We show that the estimated decay time is consistent with the result of the fitting data analysis of SXS:BBH:0305 performed by Giesler et al. \cite{Giesler:2019uxc}. Also, we estimate the source factor, $T_{22n}$, of SXS:BBH:0305 and confirm that the estimated $T_{22n}$ has the small dependence on the overtone number, contrary to $E_{22n}$. This means that at least for SXS:BBH:0305, the significance of overtones found in Ref. \cite{Giesler:2019uxc} is determined mostly by the QNEFs. In Sec. \ref{sec:conclusion}, we summarize and discuss our result that theoretically predicts the importance of the overtones and raise some intriguing issues to be studied. We set the Newton's constant to unity $G = 1$ and our computations are performed in the normalization of $2M=1$ throughout the manuscript.

\section{Excitation factors of a spinning black hole}
\label{sec:QNmode_QNEF}
\subsection{black hole perturbations and quasinormal modes}
The relaxation process of a perturbed black hole leads to ringdown radiation, which is the superposition of QN modes. We here introduce the perturbation variable of a spinning black hole and review the definition of the QN modes. The line element of the Kerr metric in the Boyer–Lindquist coordinates is
\begin{equation}
ds^2 = - \left( 1- \frac{2Mr}{\Sigma} \right) dt^2 + \frac{\Sigma}{\Delta} dr^2 + \Sigma d\theta^2 + \sin^2 \theta \left( r^2+a^2 + \frac{2M r a^2}{\Sigma} \right) d\varphi^2 - \frac{4M r a \sin^2 \theta}{\Sigma} dt d\varphi,
\end{equation}
where
\begin{align}
\Sigma &\equiv r^2 + a^2 \sin^2 \theta,\\
\Delta &\equiv r^2 -2M r + a^2 = (r-r_+) (r-r_-),\\
r_+ &\equiv M +\sqrt{M^2-a^2}, \ r_- \equiv M- \sqrt{M^2-a^2}.
\end{align}
The perturbations of the Kerr spacetime is expressed by the Newman-Penrose quantity \cite{Newman:1966ub}, $\psi_4$, which can be separated with the radial and angular components as
\begin{equation}
\psi_4 = (r-ia \cos \theta)^{-4} \int d \omega e^{-i\omega t} \sum_{l,m} e^{im\varphi} \frac{ S_{lm} (\omega, \theta)}{\sqrt{2 \pi}} R_{lm} (\omega,r),
\end{equation}
where $ R_{lm} (\omega, r)$ is the Teukolsky variable. The SWSH function $S_{lm} (\omega, \theta)$ is normalized by
\begin{equation}
\int^{\pi}_0 d\theta |S_{lm} (\omega,\theta)|^2 \sin \theta = 1.
\end{equation}
The Teukolsky variable, $R_{lm} (\omega,r)$, obeys the Teukolsky equation:
\begin{equation}
\Delta^2 \frac{d}{dr} \left( \frac{1}{\Delta} \frac{dR_{lm}(\omega, r)}{dr} \right) - V_{lm}(\omega,r) R_{lm}(\omega, r) = \tilde{T}_{lm} (\omega,r),
\label{radial_teuko}
\end{equation}
where $\tilde{T}_{lm} (\omega,r)$ is the source term and the angular momentum barrier $V_{lm}$ has the form
\begin{equation}
V_{lm} (\omega,r) \equiv - \frac{K^2 + 4i (r-M) K}{\Delta} + 8i\omega r + \lambda_{lm},
\end{equation}
where $K\equiv (r^2+a^2) \omega - ma$ and $\lambda_{lm}$ is the separation constant. The angular equation defining the SWSH function for gravitational field is \cite{Teukolsky:1973ha}
\begin{equation}
\frac{d}{du} \left[ (1-u^2) \frac{d S_{lm}}{du} \right] +\left[ 2a \omega (m+2 u) -(a\omega)^2 (1-u^2) +2 + \lambda_{lm} - \frac{(m+a \omega u)^2}{1-u^2} \right] S_{lm} = 0,
\end{equation}
where $u \equiv \cos \theta$ and $\lambda_{lm} \to l (l+1) -6$ for $a \omega \to 0$. The QN mode of a spinning black hole, $\omega = \omega_{lmn}$, is obtained by imposing the following boundary conditions to the homogeneous radial and angular equations:
\begin{align}
R_{lm} (\omega, r) \sim
\begin{cases}
\Delta^2 e^{-i k r^{\ast}} &\text{for} \ r^{\ast} \to -\infty,\\
r^3 e^{i\omega r^{\ast}} &\text{for} \ r^{\ast} \to +\infty,
\end{cases}\\
S_{lm} (\omega, \theta) \sim
\begin{cases}
(1+u)^{k_-} & \text{for} \ u \to -1,~~\\
(1-u)^{k_+} & \text{for} \ u \to +1,
\end{cases}
\label{spin_weighted_BC}
\end{align}
where $d r^{\ast}/dr \equiv (r^2+a^2)/\Delta$, $k \equiv \omega - ma/(2 M r_+)$ and $k_{\pm} \equiv |m \mp 2|/2$. The eigenvalues, $\lambda_{lm} = \lambda_{lm} (a \omega)$, is determined so that the boundary condition (\ref{spin_weighted_BC}) is satisfied for a fixed $\omega$. On the other hand, the QN modes are the zero points of the Wronskian ${\cal W}_{lm} (\omega)$ defined in the complex frequency plane:
\begin{equation}
{\cal W}_{lm} (\omega) \equiv \Delta^{-1} \left( R^{\rm (H)}_{lm} \frac{d}{dr} R^{(\infty)}_{lm} - R^{(\infty)}_{lm} \frac{d}{dr} R^{\rm (H)}_{lm} \right),
\end{equation}
where $R^{\rm (H)}$ and $R^{(\infty)}$ are the homogeneous solutions of the Teukolsky equation:
\begin{align}
R^{\rm (H)}_{lm} (\omega, r) =
\begin{cases}
A^{(\text{trans})}_{lm} (\omega) \Delta^2 e^{-i k r^{\ast}} &\text{for} \ r^{\ast} \to -\infty,\\
r^{-1} A^{(\inn)}_{lm} (\omega) e^{-i\omega r^{\ast}} + r^3 A^{(\out)}_{lm} (\omega) e^{i\omega r^{\ast}} &\text{for} \ r^{\ast} \to +\infty,
\end{cases}\\
\nonumber\\
R^{(\infty)}_{lm} (\omega, r) =
\begin{cases}
B^{(\text{in})}_{lm} (\omega) \Delta^2 e^{-i k r^{\ast}} + B^{(\text{out})}_{lm} (\omega) e^{+i k r^{\ast}} &~~\text{for} \ r^{\ast} \to -\infty,\\
r^3 B^{\rm (trans)}_{lm} (\omega) e^{i\omega r^{\ast}} &~~\text{for} \ r^{\ast} \to +\infty.
\end{cases}
\end{align}
The Wronskian reduces to 
\begin{equation}
{\cal W}_{lm} (\omega) = 2 i \omega A^{(\inn)}_{lm} (\omega),
\end{equation}
which vanishes at $\omega = \omega_{lmn}$
\begin{equation}
\displaystyle
\lim_{\omega \to \omega_{lmn}} {\cal W}_{lm} (\omega) = \lim_{\omega \to \omega_{lmn}} 2 i \omega (d A^{(\inn)}_{lm}/d \omega) (\omega - \omega_{lmn}) = 0.
\end{equation}

\subsection{definition}
A spinning black hole has the infinite number of discretized QN modes, and a ringdown signal can be represented by the superposition of QN modes. Of course, not all of them are significantly excited in a perturbed black hole, and it may be natural to ask how we can quantify the ease of excitation for each QN mode. This is crucial, for example, to model the waveform of rigndown signal emitted by a spinning black hole. It is known that the QNEF, $E_{lmn}$, is very the quantity that quantifies the ease of excitation of QN modes and is independent of the source of perturbation \cite{Sun:1988tz,Leaver:1986gd,Andersson:1995zk,Glampedakis:2001js,Glampedakis:2003dn,Berti:2006wq,Zhang:2013ksa}. Here we review the definition of QNEF and slightly modify its original definition \cite{Zhang:2013ksa}, in which the QNEF depends on the perturbation variables. The standard perturbation variables are the Teukolsky \cite{Teukolsky:1973ha}, Sasaki-Nakamura \cite{Sasaki:1981sx}, and Chandrasekhar-Detweiler \cite{Chandrasekhar:1976zz} variables for the Kerr spacetime. To quantify the importance of overtones measured in strain, we here introduce the QNEF slightly modified based on the strain amplitude $h = h_{+} + i h_{\times}$ that has the form of
\begin{align}
\displaystyle
h= \lim_{r^{\ast} \to \infty} h_{+} +i h_{\times}
= - \frac{2}{r} \frac{e^{im \phi}}{\sqrt{2 \pi}} \int^{\infty}_{-\infty} d \omega \sum_{lm} \frac{e^{i \omega (r^{\ast} - t+t_s)}}{\omega^2} S_{lm} (\omega, \theta) \frac{A_{lm}^{(\out)}}{2i\omega A_{lm}^{(\inn)}} \int_{r_+}^{\infty} dr' \frac{\tilde{T}_{lm} (\omega, r') R^{\rm (H)}_{lm} (\omega, r')}{A_{lm}^{(\out)} \Delta^2 (r')},
\label{metric_per}
\end{align}
where $t_s$ is the start time of ringdown. The integration with respect to $\omega$ in Eq. (\ref{metric_per}) can be replaced by the summation of the residues at the QN frequencies, $\omega = \omega_{lmn}$, since $A_{lm}^{(\inn)} (\omega) \sim (\omega-\omega_{lmn})$ for $\omega \sim \omega_{lmn}$. Therefore, the metric perturbations $h (t,r)$ can be written as
\begin{align}
h(t,r) = - \frac{2}{r} \frac{e^{im\phi}}{\sqrt{2 \pi}} (2\pi i) \sum_{lmn} E_{lmn} T_{lmn} S_{lmn} (\theta) e^{i \omega_{lmn} (r^{\ast} - t+t_s)},
\label{metric_perturb_ET}
\end{align}
where 
\begin{align}
E_{lmn} &\equiv \frac{A_{lm}^{(\out)} (\omega_{lmn})}{2 i \omega_{lmn}^3} \left( \frac{dA_{lm}^{(\inn)}}{d\omega} \right)_{\omega = \omega_{lmn}}^{-1},
\label{EF_formula}\\
T_{lmn} &\equiv \int_{r_+}^{\infty} dr' \frac{\tilde{T}_{lm} (\omega_{lmn}, r') R^{\rm (H)}_{lm} (\omega_{lmn}, r')}{A^{(\out)}_{lm} (\omega_{lmn}) \Delta^2 (r')},\\
S_{lmn} &\equiv S_{lm} (\omega_{lmn},\theta).
\end{align}
Our definition of the excitation factor Eq. (\ref{EF_formula}) is suited to quantify the ease of excitation in strain. Note that the QNEF $E_{lmn}$ is independent of the source of perturbation and is determined only by the mass and spin of a black hole. On the other hand, the factor $T_{lmn}$ is determined by the source term $\tilde{T}_{lm}(\omega,r)$ in Eq. (\ref{radial_teuko}), which means that $T_{lmn}$ includes the information of the source of perturbation. The SWSH factor $S_{lmn}$ is also the quantity independent of the source of perturbation and determines the amplitude of each QN mode. However, its dependence on the overtone number, $n$, is weaker than the $n$-dependence of QNEF as is shown in Appendix \ref{app:SWSHF}. Therefore, we conclude that the $n$-dependence of the ease of excitation of QN modes is mostly determined by $E_{lmn}$ rather than $S_{lmn}$. Also, in Sec. \ref{sec:decaytime}, we will show the $n$-dependence of $T_{22n}$ for SXS:BBH:0305, which is also weaker than that of $E_{22n}$.

\subsection{Overview of our methodology to compute the excitation factors}
\label{sec:methodology}
In this section, we briefly overview how we compute the QN frequencies and QNEFs, and all other details are provided in the Appendix \ref{app:QN frequencies} and \ref{app:QNEFs}. The main procedures of our computation for the $n$-th QNEF can be summarized as follows:
\begin{itemize}
\item[Step 1] Compute the $n$-th QN frequency, $\omega_{lmn} (\Lambda)$, of a Kerr-de Sitter (KdS) black hole by using the general Heun function \cite{Suzuki:1998vy}. The Mathematica function, {\sffamily HeunG}, is available after the version 12.1 (c.f. \cite{Hatsuda:2020sbn}).
\item[Step 2] Repeat the first procedure in the range of $\Lambda = 0.002 - 0.02$ ($M=1/2$) with a step size of $N=41$ iterations.
\item[Step 3] Compute the corresponding values of the $n$-th QNEF, $E_{lmn} (\Lambda)$, with the same range and step size of $\Lambda$ as in Step 2.
\item[Step 4] Extrapolate the values of the $n$-th QNEF to the value of $\Lambda = 0$ and obtain the $n$-th QNEF of the Kerr black hole.
\end{itemize}
The Teukolsky equation for the KdS black hole reduces to the Heun equation \cite{Suzuki:1998vy}, and hence, it has an analytic solution represented by the general Heun function. This allows us to compute the QN frequencies for $\Lambda > 0$ with high precision by using Mathematica 12.1 or later versions. The computation of QN frequencies $\omega_{lmn}$ for a Kerr black hole by extrapolating the QN frequencies of the KdS spacetime to those of $\Lambda = 0$ by using Mathematica was originally suggested in \cite{Hatsuda:2020sbn}. We here extend it to the computation of QNEFs and QN frequencies including overtones. The obtained QN frequencies are in agreement with the catalog provided in \cite{Berti:cite,grit:cite}. In the Appendix \ref{app:QNtable}, our result of the QN frequencies, $\omega_{22n}$, up to $n=20$ are shown for the various spin parameters. The $n$-th QNEF, $E_{lmn}$, is obtained by computing the derivative of $A_{lm}^{(\inn)}$ with respect to $\omega$ near the QN frequencies\footnote{We numerically compute $d A_{lm}^{(\inn)}/d\omega$ by the finite difference method as $(A_{lm}^{(\inn)} (\omega + \delta \omega) - A_{lm}^{(\inn)} (\omega))/\delta \omega$, and we take $\delta \omega = 10^{-9}$ for which the value of QNEF converges. We also repeat the computation with $\delta \omega = i 10^{-9}$ and confirm that the values are in agreement with each other.} (see (\ref{EF_formula})) and by performing proper normalization (Step 3). For the details of the normalization to compute the QNEF, see Appendix \ref{app:QNEFs}. Extrapolating $E_{lmn} (\Lambda)$ to the value of $\Lambda =0$, one can obtain the QNEFs for the Kerr spacetime (Step 4). We set a fitting function as
\begin{equation}\displaystyle
E_{lmn} (\Lambda) \simeq \sum_{k=0}^{N-1} c_k \left( \frac{1}{\ell} \right)^{k},
\label{excitation_fitting}
\end{equation}
where $\ell \equiv \sqrt{3/\Lambda}$. The precision of Step 4 can be improved by increasing the iteration number, $N$, in Step 2. The numerical convergence of the value of $E_{lmn} (\Lambda)$ and comparison with the result in Ref. \cite{Zhang:2013ksa} is shown in Appendix \ref{app:validity}. 

Although our procedure makes the computation of QNEFs up to higher overtones possible with high accuracy, we find that for lower spins ($j\lesssim 0.2$), the fitting function (\ref{excitation_fitting}) does not work because $\omega_{lmn} (\Lambda)$ and $E_{lmn} (\Lambda)$ has a modulation with respect to $\Lambda$ for slow rotations. In this paper, the main motivation is to quantify the importance of overtones for ringing remnant black holes resulting from BBH mergers, whose remnant spin parameters are typically in $0.6 \lesssim j \lesssim 0.9$, and hence, we can take the advantage of this procedure.

\subsection{Excitation factors for Kerr black holes}
\label{sec:QNEF_result}
\begin{figure}[H]
  \begin{center}
    \includegraphics[keepaspectratio=true,height=60mm]{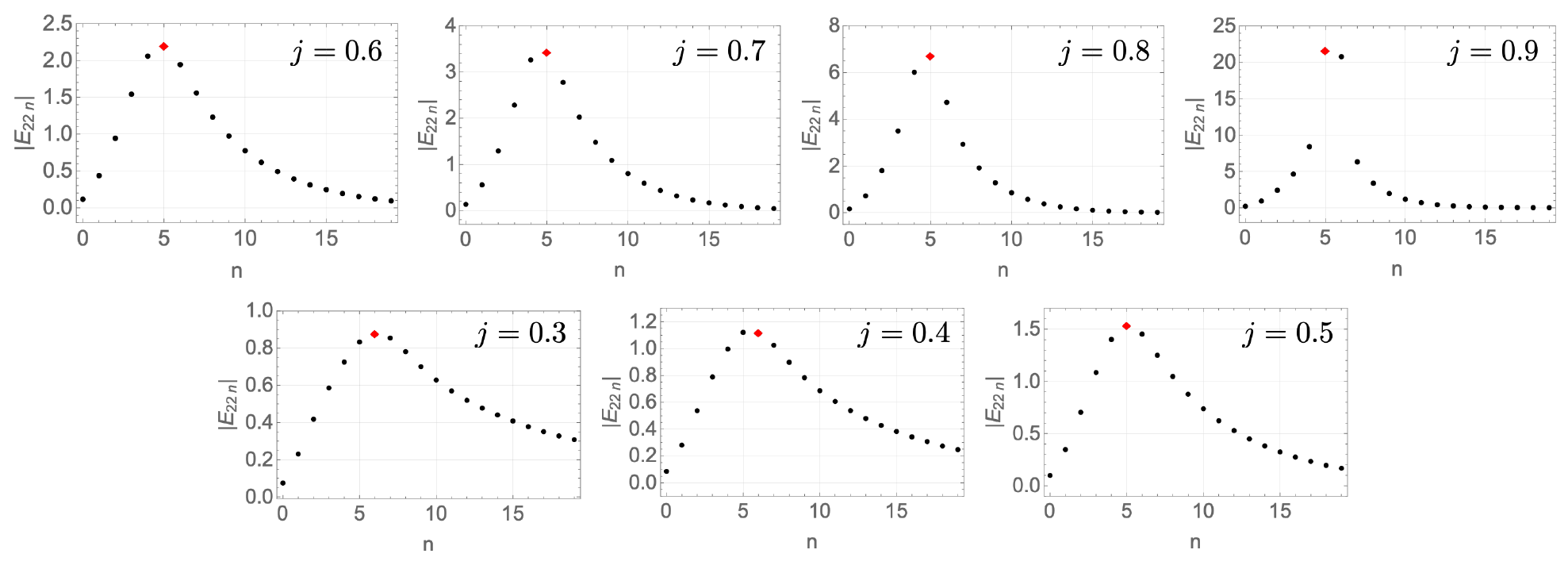}
  \end{center}
\caption{The QNEFs for $j=0.3$, $0.4$, $0.5$, $0.6$, $0.7$, $0.8$, and $0.9$ with $l=m=2$. The maximum value of QNEF is indicated by red points.
}
\label{ef_high_spin}
\end{figure}

The obtained values of QNEFs, $E_{22n}$, for $0.3 \leq j \leq 0.9$ and $n =0, 1, 2, \cdots, 20$ are shown in FIG. \ref{ef_high_spin} and the data of their absolute values and arguments are shown in Table \ref{table_EF}. 
%%%%%%%%%%%%%%%%%%%%%%%%%
%%%%%%%%%%%%%%%%%%%%%%%%%
%%%%%%%%%%%%%%%%%%%%%%%%%
\begin{figure}[H]
  \begin{center}
    \includegraphics[keepaspectratio=true,height=60mm]{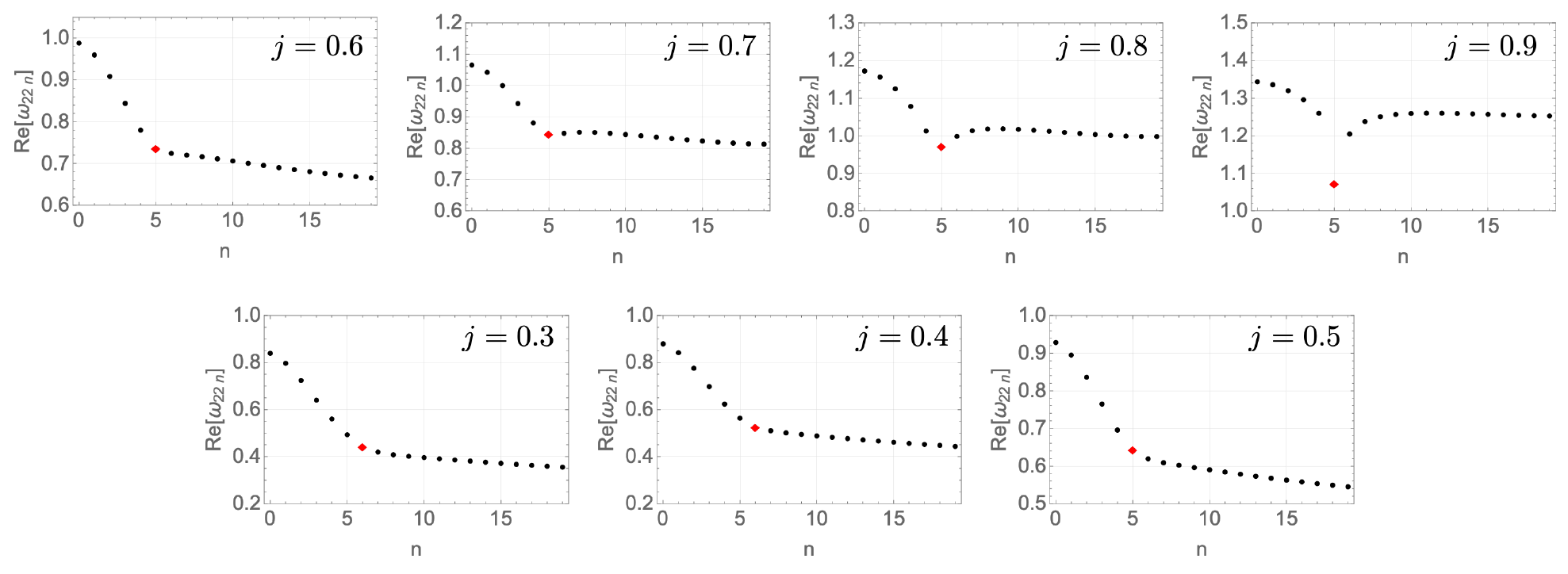}
  \end{center}
\caption{The real part of QN frequencies for $j=0.3$, $0.4$, $0.5$, $0.6$, $0.7$, $0.8$, and $0.9$ with $l=m=2$. The red points indicate the overtone which has the maximum value of QNEF.
}
\label{f_corner}
\end{figure}
%%%%%%%%%%%%%%%%%%%%%%%%%
We find that the peak of the QNEF is at $n=5$ for $0.5 \lesssim j \lesssim 0.9$ which covers the typical spin parameters of remnant black holes resulting from BBH mergers. For lower spin parameters ($j \lesssim 0.4$), on the other hand, the peak comes at $n=6$. Based on the values of QNEFs obtained from our computation, we find that the 4th, 5th, and 6th QN modes are the top three important overtones. This result is consistent with the previous work \cite{Giesler:2019uxc} in which the excitation of overtones was investigated from the fitting data analysis of waveforms in the SXS collaboration catalog \cite{Boyle:2019kee}.

One can see that the QN mode which has the maximum value of $E_{22n}$ is at the {\it corner} of the QN-frequency curve (see FIG. \ref{f_corner}). The correspondence of the peak in QNEF and the corner in the QN-frequency curve is interesting and this implies that there may be a nontrivial correlation between the behaviors of $\omega_{lmn}$ and $E_{lmn}$.

\begin{figure}[h]
  \begin{center}
    \includegraphics[keepaspectratio=true,height=86mm]{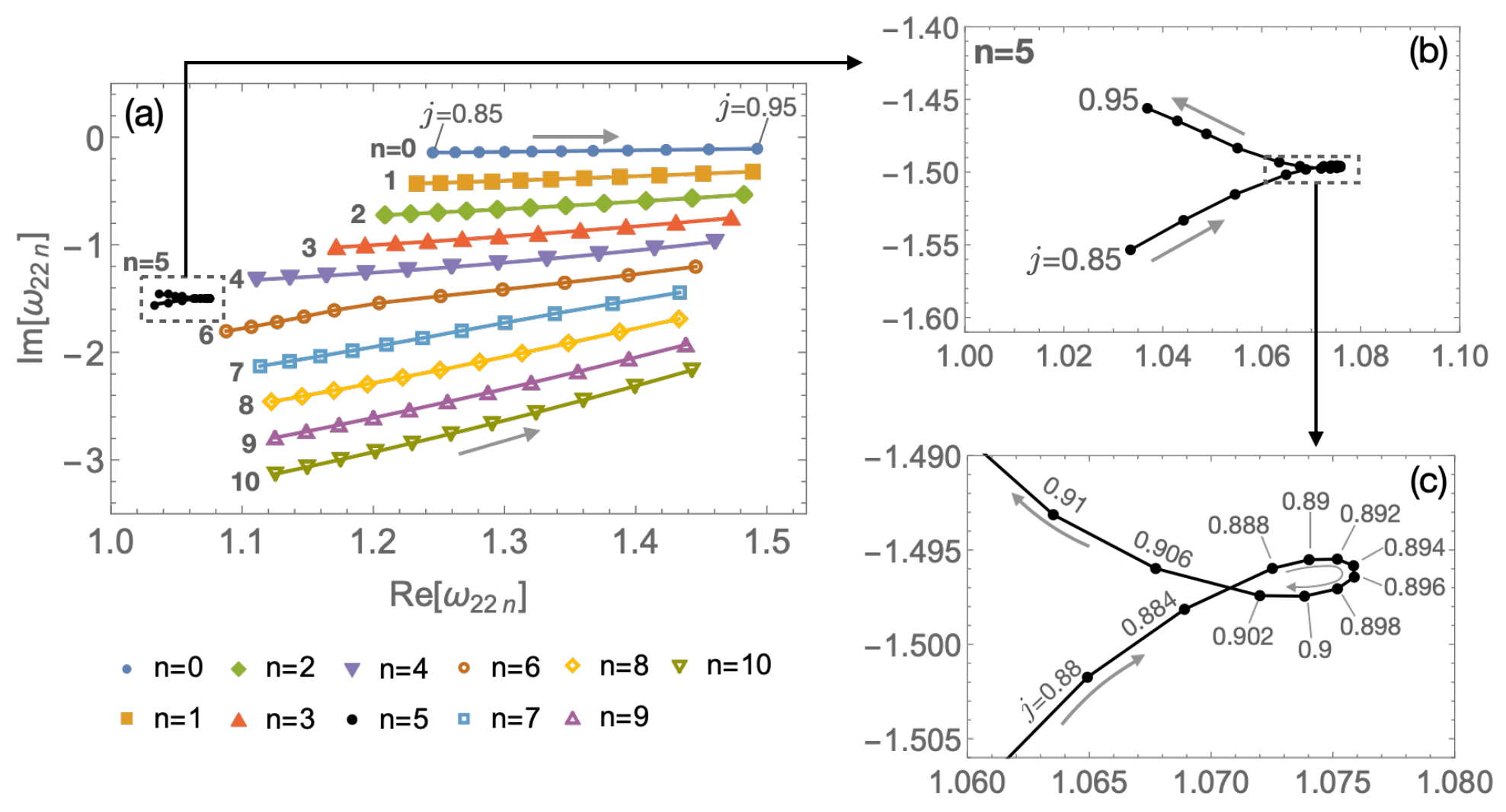}
  \end{center}
\caption{(a) Plot of the QN frequencies, $\omega_{22n}$, for $n=0, 1, 2,\cdots, 10$ and $0.85 \leq j \leq 0.95$. The gray arrows indicate the direction for which the spin parameter increases. The 5th QN mode has anomalous behaviour and it has the peak value of $\text{Re} (\omega_{22n})$ around $j = 0.9$ ((b) and (c)).
}
\label{QN mode_5th}
\end{figure}

\begin{figure}[h]
  \begin{center}
    \includegraphics[keepaspectratio=true,height=55mm]{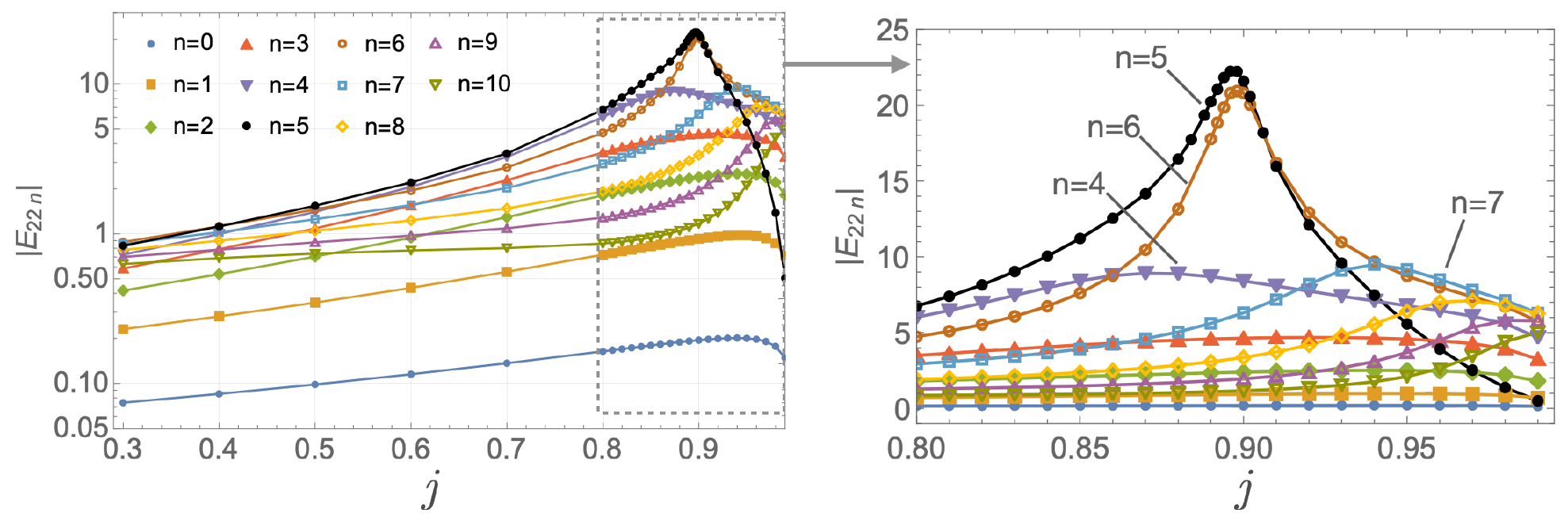}
  \end{center}
\caption{Plot of the QNEF, $E_{22n} (j)$, for $n=0, 1, 2, \cdots, 10$. The 5th QNEF has its peak at $j \simeq 0.9$ and suppressed in the near-extremal situation.
}
\label{QNEF_5th}
\end{figure}

\begin{figure}[h]
  \begin{center}
    \includegraphics[keepaspectratio=true,height=40mm]{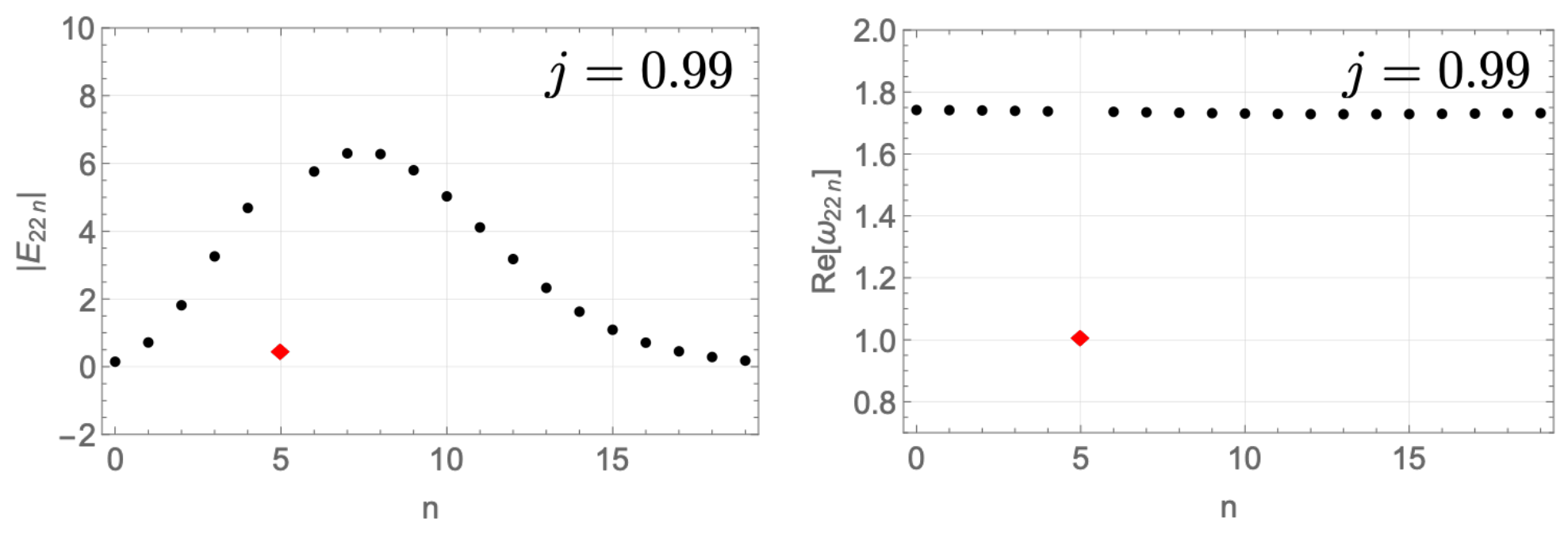}
  \end{center}
\caption{QNEFs (left) and the real part of the QN frequencies (right) for the near-extremal limit $j=0.99$ with $l=m=2$.
}
\label{QN_mode_099}
\end{figure}

The closer to the extremal limit the black hole is, the sharper the QN-frequency curve at the 5th QN mode is. In the extremal limit, the 5th QN mode is isolated from the other QN modes (FIG. \ref{QN mode_5th}), and this anomalous behaviour of $\omega_{225}$ was originally found by Onozawa \cite{Onozawa:1996ux}. The 5th QN frequency has a loop trajectory around $j = 0.9$ (see FIG. \ref{QN mode_5th}-(c)), where the 5th QNEF has the maximum absolute value as is shown in FIG. \ref{QNEF_5th}. The anomalous behaviour near the extremal limit can be seen in the QNEF as well. The closer to the extremal limit the black hole is, the more suppressed the 5th QNEF is (see FIG. \ref{QNEF_5th} and \ref{QN_mode_099}). Although the anomalous behaviour of the 5th QN mode for the near-extremal situation is interesting, as far as I know, it is still an open question why the 5th QN mode is so special.
%%%%%%%%%%%%%%%%%%%%%%%%%
%%%%%%%%%%%%%%%%%%%%%%%%%
%%%%%%%%%%%%%%%%%%%%%%%%%
%%%%%%%%%%%%%%%%%%%%%%%%%
%%%%%%%%%%%%%%%%%%%%%%%%%
%%%%%%%%%%%%%%%%%%%%%%%%%

\section{Excitation of overtones by a particle plunging into a black hole}
\label{sec:plungeGW}
In this section, we compute the GW waveform sourced by a particle plunging into a spinning black hole to see how important the overtones are in an independent way from the computation of the QNEF. Especially, for intermediate and high spin parameters, the excitation factors of the 4th, 5th, and 6th overtones are the three highest modes. For the near-extremal case ($j \sim 0.99$), on the other hand, the 5th QNEF is suppressed, and the higher overtones $6 \leq n \leq 9$ have relatively larger absolute values of QNEFs. We here check how such behaviour of the QNEFs are reflected in GW emission induced by a particle plunging into a spinning black hole. To this end, we compute the GW waveform and extract the amplitude of each QN mode, $\tilde{C}_{22n}$, from the waveform.

As the first step, we will solve the Sasaki-Nakamura equation with the source term of a plunging particle whose trajectory is on the equatorial plane $(\theta = \pi/2)$ \cite{Kojima:1984cj,Nakamura:1987zz}. Also, we assume that an observer to detect the emitted GW signal is on the equatorial plane and at a distant region $r^{\ast} \gg 1$. Then we fit the QN modes of $l=m=2$ and $n=0, 1, 2, \cdots, 7$ to the obtained GW waveform and extract the amplitude of each QN mode, which is given by the product of $E_{22n}$, $T_{22n}$, and $S_{22n}$. In this work, we set the fit start time, $t_{\rm fit}$, to the moment when the plunging particle starts to follow the null geodesics near the black hole horizon (see Appendix \ref{app:fit}). After the moment, the particle is absorbed by the black hole and the system relaxes to an axial-symmetric Kerr black hole.

\begin{table}

\begin{center}
\begin{tabular}{c|cc|cc|cc|cc}
\firsthline
overtone number &\multicolumn{2}{c|}{$j=0.7$} &\multicolumn{2}{c|}{$j=0.8$}&\multicolumn{2}{c|}{$j=0.9$}&\multicolumn{2}{c}{$j=0.99$} \\
\cline{2-9}
$n$    & $|E_{22n}|$ & arg$(E_{22n})$ & $|E_{22n}|$ & arg$(E_{22n})$ & $|E_{22n}|$ & arg$(E_{22n})$& $|E_{22n}|$ & arg$(E_{22n})$ \\
\hline
$0$      & 0.136 & 0.879 & 0.164 & 1.17 & 0.194 & 1.62 & 0.148 & 2.71 \\
$1$      & 0.557 & $-$1.31 & 0.725 & $-$0.917 & 0.927 & $-$0.292 & 0.716 & 1.15 \\
$2$      & 1.29 & 2.64 & 1.80 & $-$3.10 & 2.39 & $-$2.22 & 1.81 & $-$0.285 \\
$3$      & 2.28 & 0.155 & 3.50 & 0.885 & 4.63 & 2.12 & 3.25 & $-$1.62\\
$4$      & 3.26 & $-$2.56 & 6.01 & $-$1.68 & 8.40 & 0.250 & 4.68 & $-$2.88 \\
$5$      & 3.44 & 0.825 & 6.75 & 1.70 & 21.6 & 2.60 & 0.505 & 0.747 \\
$6$      & 2.77 & $-$2.04 & 4.72 & $-$1.13 & 20.8 & $-$0.605 & 5.76 & 2.22 \\
$7$      & 2.02 & 1.49 & 2.93 & 2.62 & 6.31 & $-$2.01 & 6.30 & 1.10 \\
$8$      & 1.48 & $-$1.20 & 1.92 & 0.170 & 3.36 & 2.34 & 6.28 & 0.0509 \\
$9$      & 1.09 & 2.40 & 1.28 & $-$2.26 & 1.96 & 0.362 & 5.80 & $-$0.94 \\
$10$     & 0.802 & $-$0.278 & 0.860 & 1.58 & 1.17 & $-$1.63 & 5.03 & $-$1.87 \\
$11$     & 0.591 & $-$2.96 & 0.576 & $-$0.852 & 0.701 & 2.66 & 4.11 & $-$2.76 \\
$12$     & 0.435 & 0.640 & 0.384 & 2.99 & 0.419 & 0.657 & 3.18 & 2.68 \\
$13$     & 0.318 & $-$2.04 & 0.254 & 0.550 & 0.249 & $-$1.34 & 2.33 & 1.87 \\
$14$     & 0.231 & 1.54 & 0.167 & $-$1.89 & 0.147 & 2.94 & 1.62 & 1.06 \\
$15$     & 0.168 & $-$1.14 & 0.109 & 1.94 & 0.0869 & 0.936 & 1.09 & 0.268 \\
$16$     & 0.121 & 2.44 & 0.0714 & $-$0.500 & 0.0508 & $-$1.06 & 0.710 & $-$0.534 \\
$17$     & 0.0873 & $-$0.249 & 0.0461 & $-$2.95 & 0.0295 & $-$3.07 & 0.453 & $-$1.34 \\
$18$     & 0.0624 & $-$2.94 & 0.0296 & 0.883 & 0.0171 & 1.21 & 0.285 & $-$2.16 \\
$19$     & 0.0443 & 0.635 & 0.0189 & $-$1.57 & 0.00983 & $-$0.792 & 0.178 & $-$2.98 \\
$20$     & 0.0313 & $-$2.07 & 0.0120 & 2.24 & 0.00563 & $-$2.80 & 0.110 & 2.48 \\
\lasthline
\end{tabular}
\end{center}

\begin{center}
\begin{tabular}{c|cc|cc|cc|cc}
\firsthline
overtone number &\multicolumn{2}{c|}{$j=0.3$} &\multicolumn{2}{c|}{$j=0.4$}&\multicolumn{2}{c|}{$j=0.5$}&\multicolumn{2}{c}{$j=0.6$} \\
\cline{2-9}
$n$    & $|E_{22n}|$ & arg$(E_{22n})$ & $|E_{22n}|$ & arg$(E_{22n})$ & $|E_{22n}|$ & arg$(E_{22n})$& $|E_{22n}|$ & arg$(E_{22n})$ \\
\hline
$0$      & 0.0743 & 0.210 & 0.0850 & 0.340 & 0.0983 & 0.488 & 0.115 & 0.662   \\
$1$      & 0.230 & $-$2.14 & 0.280 & $-$1.99 & 0.346 & $-$1.81 & 0.435 & $-$1.59 \\
$2$      & 0.417 & 1.61 & 0.537 & 1.78 & 0.703 & 2.00 & 0.942 & 2.27 \\
$3$      & 0.585 & $-$1.08 & 0.788 & $-$0.885 & 1.08 & $-$0.635 & 1.54 & $-$0.307 \\
$4$      & 0.725 & 2.32 & 0.996 & 2.54 & 1.40 & 2.82 & 2.05 & $-$3.09 \\
$5$      & 0.832 & $-$0.696 & 1.12 & $-$0.462 & 1.53 & $-$0.152 & 2.20 & 0.258 \\
$6$      & 0.881 & 2.46 & 1.12 & 2.73 & 1.45 & 3.10 & 1.94 & $-$2.69 \\
$7$      & 0.854 & $-$0.702 & 1.02 & $-$0.360 & 1.25 & 0.0932 & 1.55 & 0.689 \\
$8$      & 0.781 & 2.43 & 0.898 & 2.85 & 1.04 & $-$2.87 & 1.23 & $-$2.16 \\
$9$      & 0.700 & $-$0.678 & 0.783 & $-$0.179 & 0.877 & 0.454 & 0.973 & 1.28 \\
$10$      & 0.628 & 2.51 & 0.686 & 3.07 & 0.737 & $-$2.49 & 0.774 & $-$1.55 \\
$11$      & 0.569 & $-$0.567 & 0.605 & 0.0565 & 0.623 & 0.847 & 0.616 & 1.89 \\
$12$      & 0.519 & 2.64 & 0.537 & $-$2.96 & 0.528 & $-$2.10 & 0.491 & $-$0.950\\
$13$      & 0.477 & $-$0.429& 0.478 & 0.297 & 0.448 & 1.23 & 0.390 & 2.49 \\
$14$      & 0.440 & 2.78 & 0.427 & $-$2.72 & 0.381 & $-$1.71 & 0.310 & $-$0.356 \\
$15$      & 0.408 & $-$0.292 & 0.382 & 0.533 & 0.323 & 1.61 & 0.245 & 3.07 \\
$16$      & 0.377 & 2.91 & 0.342 & $-$2.49 & 0.274 & $-$1.34 & 0.193 & 0.228 \\
$17$      & 0.350 & $-$0.159 & 0.307 & 0.763 & 0.233 & 1.98 & 0.152 & $-$2.62 \\
$18$      & 0.327 & 3.04 & 0.273 & $-$2.26 & 0.195 & $-$0.970 & 0.119 & 0.805 \\
$19$      & 0.307 & $-$0.0196 & 0.245 & 0.989 & 0.167 & 2.34 & 0.0934 & $-$2.04 \\
$20$      & 0.284 & $-$3.11 & 0.219 & $-$2.04 & 0.140 & $-$0.593 & 0.0726 & 1.37 \\
\lasthline
\end{tabular}
\end{center}
\caption{The absolute values and arguments of the QNEFs with $n=0, 1, 2, \cdots, 20$ and $l=m=2$ for a Kerr black hole. The spin parameter is set to $j=0.3$, $0.4$, $0.5$, $0.6$, $0.7$, $0.8$, $0.9$, and $0.99$.}
\label{table_EF}
\end{table}

\subsection{GW emission induced by a particle plunging into a black hole}
The trajectory of a particle plunging into a black hole on the equatorial plane ($\theta = \pi/2$) is governed by the following equations \cite{Kojima:1984cj,Nakamura:1987zz}
\begin{align}
r^2 \frac{dt}{d\tau} &= -a (a-L) + \frac{r^2+a^2}{\Delta (r)} P(r),\\
r^2 \frac{dr}{d\tau} &= - \sqrt{Q(r)},\\
r^2 \frac{d\varphi}{d\tau} &= -(a-L) + \frac{a}{\Delta (r)} P(r),
\end{align}
where $\tau$ is the proper time, $L$ is the orbital angular momentum of the particle, and
\begin{align}
Q(r) &\equiv 2M r^3 - L^2 r^2 + 2M r (a-L)^2,\\
P(r) &\equiv r^2 + a^2 -aL.
\end{align}
The source term of the plunging particle in the Sasaki-Nakamura equation is \cite{Kojima:1984cj,Nakamura:1987zz}
\begin{equation}
\tilde{T}^{\rm (SN)}_{lm} = \frac{\gamma \Delta \mu \tilde{W}}{(r^2 +a^2)^{3/2} r^2} e^{ -i k r^{\ast}},
\label{SN_source}
\end{equation}
where $\gamma \equiv \lambda (\lambda+2) -12i M \omega-12a \omega (a\omega-m)$ and the explicit form of $\tilde{W}$ is given in the Appendix \ref{app:source}. The strain amplitude in the far region ($r^{\ast} \to \infty$) is
\begin{align}
\displaystyle
h_{+} + i h_{\times} &= - \frac{2}{r} \frac{1}{\sqrt{2 \pi}} \int^{\infty}_{- \infty} d \omega \sum_{lm} \frac{e^{i\omega (r^{\ast} -t)}}{\omega^2} R^{\rm(out)}_{lm} (\omega) S_{lm} (\omega, \theta) e^{im \phi},\\
R^{\rm(out)}_{lm} &= - \frac{4 \omega^2 X_{lm} (\omega)}{\lambda (\lambda+2) -12i M \omega -12 a^2 \omega^2} r^3 e^{\i \omega r^{\ast}},\\
X_{lm} &\simeq \int dr' \frac{\tilde{T}^{\rm (SN)}_{lm} (r', \omega) X^{\rm (in)}_{lm} (r',\omega)}{2i\omega B (\omega)} \ \text{for} \ r \to \infty,
\end{align}
where $X^{(\inn)}_{lm}$ is the homogeneous solution of the Sasaki-Nakamura equation that is purely ingoing at the black hole horizon
\begin{align}
X^{(\inn)}_{lm} (r, \omega) =
\begin{cases}
A (\omega) e^{i\omega r^{\ast}} + B (\omega) e^{-i \omega r^{\ast}} & (r \to \infty),\\
e^{-i k r^{\ast}} & (r^{\ast} \to - \infty).
\end{cases}
\end{align}
%%%%%%%%%%%%%%%%%%%%%%%%%
\begin{figure}[t]
  \begin{center}
    \includegraphics[keepaspectratio=true,height=95mm]{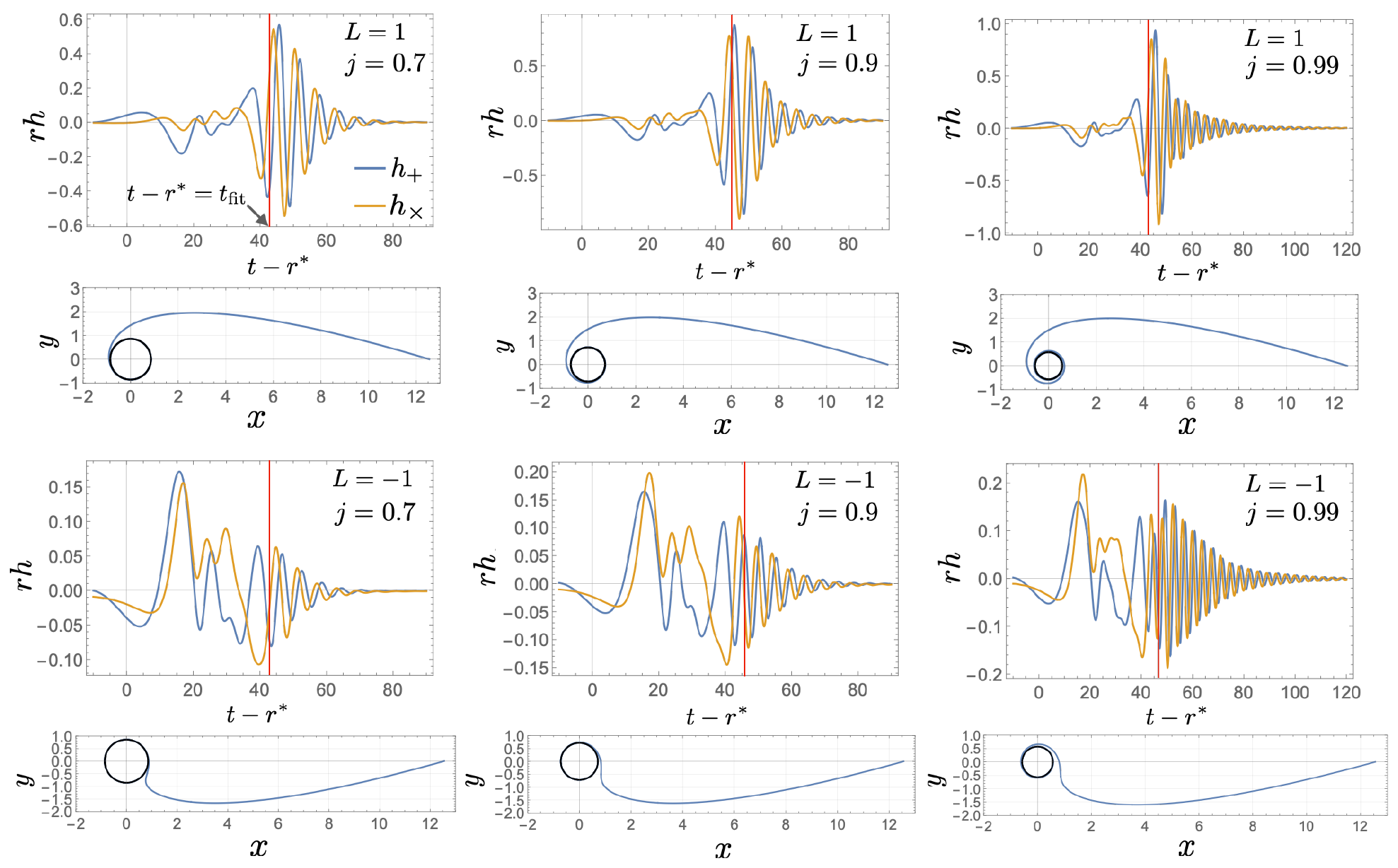}
  \end{center}
\caption{
Strain amplitudes, $r h(t-r^{\ast})$, and the trajectories of a particle on the equatorial plane ($\theta = \pi/2$). The mass of the particle is set to $\mu=1$, and the spin parameter is set to $j=0.7$, $0.9$, and $0.99$ with a co-rotating ($L=1$) and counter-rotating cases ($L=-1$). The red lines indicate the fit start time, $t_{\rm fit}$, and the extracted amplitudes are shown in FIG. \ref{ampls}.
}
\label{waveforms}
\end{figure}
%%%%%%%%%%%%%%%%%%%%%%%%%
%%%%%%%%%%%%%%%%%%%%%%%%%
\begin{figure}[t]
  \begin{center}
    \includegraphics[keepaspectratio=true,height=65mm]{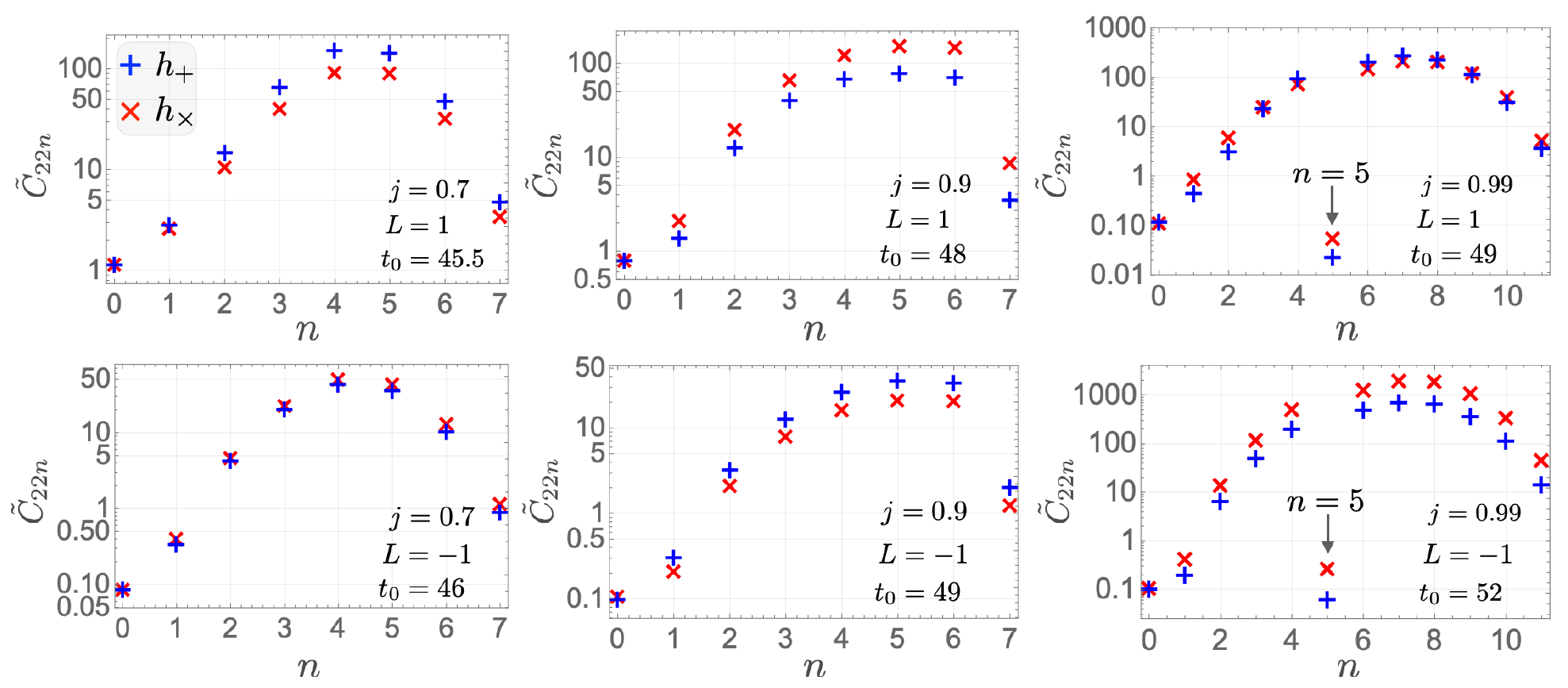}
  \end{center}
\caption{Amplitudes for each QN modes ($n=0, 1, 2, \cdots, 7$) obtained by fitting the QN modes to the waveforms shown in FIG. \ref{waveforms}.
}
\label{ampls}
\end{figure}
%%%%%%%%%%%%%%%%%%%%%%%%%
We numerically solve the Sasaki-Nakamura equation and compute the strain amplitude with the source term (\ref{SN_source}). The obtained waveforms are shown in FIG. \ref{waveforms} along with the trajectories of the point particle.
In the next subsection, we will perform the fitting analysis of the QN frequencies of $n=0, 1, 2, \cdots, 7$ to extract the amplitudes, $\tilde{C}_{22n}$, and to show which overtones are highly excited.

\subsection{Excitation of overtones}
\label{sec:fit_overtones}
We here fit the QN modes\footnote{We also subtract the ringdown tail at late time as necessary.} to the obtained waveforms shown in FIG. \ref{waveforms}, and the fitting function is
\begin{equation}
\displaystyle
h_{22} \simeq \frac{1}{r} \sum_{n=0}^{n_{\rm max}} \tilde{C}_{22n} e^{\text{Im}(\omega_{22n}) (t-r^{\ast}-t_0)}\exp \left[ \text{Re}(\omega_{22n}) (t-r^{\ast}-t_0) + i \delta_{n} \right],
\end{equation}
where $\tilde{C}_{22n}$ and $\delta_n$ are fitting parameters, and $t_0$ is a reference time. We take $n_{\rm max} = 7$ for $j=0.7$ and $0.9$ and $n_{\rm max} = 11$ for $j=0.99$. We use a Mathematica function {\sffamily Findfit} with {\sffamily MaxIterations $\to$ 100000} to fit the waveforms, and the fit start time is set to the moment when the particle starts to follow the null geodesics, which may be regarded as the absorption of the particle by the black hole horizon (see Appendix \ref{app:fit}). In FIG. \ref{ampls}, the fitted values of $\tilde{C}_{22n}$ are shown and one can see that the 4th and 5th QN modes are highly excited for $j=0.7$. For $j = 0.9$, the 5th and 6th QN modes are highly excited. In the near-extremal case $j=0.99$, on the other hand, the 5th overtone is suppressed (see FIG. \ref{QN_mode_099}) and the 7th and 8th overtones are highly excited. Also, we find that the excitation or suppression of the overtones is insensitive to the initial condition of the plunging particle, e.g., corotating and counter rotating particles. What we extracted from the waveforms is not $E_{22n}$ but $\tilde{C}_{22n} = E_{22n} T_{22n} S_{22n}$, and this fitting analysis is totally independent of the computation of $E_{22n}$. Nevertheless, the extracted amplitudes are consistent with our result of the direct computation of QNEF shown in Sec. \ref{sec:QNEF_result}. In general, the $n$-dependence of $\tilde{C}_{lmn}$ matches with that of $E_{lmn}$ when
\begin{equation}
|T_{lmn}|/|T_{lm0}| \sim {\cal O} (1) \ \text{and} \ |S_{lmn}|/|S_{lm0}| \sim {\cal O} (1).
\label{T_S_sim_1}
\end{equation}
Indeed, in our case, those conditions are satisfied as is shown in Table \ref{table_TS_sim_1}. In the latter part of the next section, we will revisit the result of the fitting data analysis of the GW150914-like numerical relativity waveform SXS:BBH:0305 done in Ref. \cite{Giesler:2019uxc} to see the consistency between our result of $E_{22n}$ and the ringdown signal of the BBH merger. Also, we will check (\ref{T_S_sim_1}) in the situation.
\begin{table}[H]
\begin{center}
\begin{tabular}{c c c c c c c c c} 
 \hline
 $n$ &0 & 1 & 2 & 3 & 4 & 5 & 6 & 7  \\ [0.5ex] 
 \hline\hline
  $|E_{22n}|/|E_{220}|$ & 1 & 4.09 & 9.47 & 16.7 & 23.9 & 25.3 & 20.3 & 14.8 \\ 
 \hline
   $|T_{22n}|/|T_{220}|$ & 1 & 0.563 & 0.962 & 2.06 & 3.22 & 2.97 & 1.30 & 0.188 \\ 
 \hline
 $|S_{22n}|/|S_{220}|$ & 1 & 1.01 & 1.02 & 1.03 & 1.05 & 1.06 & 1.07 & 1.08 \\ 
 \hline
\end{tabular} \ \ \ \ 
\begin{tabular}{c c c c c c c c c} 
 \hline
 $n$ &0 & 1 & 2 & 3 & 4 & 5 & 6 & 7  \\ [0.5ex] 
 \hline\hline
  $|E_{22n}|/|E_{220}|$ & 1 & 4.77 & 12.3 & 23.8 & 43.2 & 112 & 107 & 32.4 \\ 
 \hline
   $|T_{22n}|/|T_{220}|$ & 1 & 0.548 & 1.95 & 3.35 & 3.36 & 1.57 & 1.56 & 0.296 \\ 
 \hline
 $|S_{22n}|/|S_{220}|$ & 1 & 1.01 & 1.02 & 1.04 & 1.07 & 1.09 & 1.11 & 1.13 \\ 
 \hline
\end{tabular}
\end{center}
\caption{The $n$-dependence of the QNEF, source factor, and SWSH factor for the signal of $h_{\times}$ induced by the plunging particle ($L=1$) into the black hole of $j=0.7$ (left) and $0.9$ (right).}
\label{table_TS_sim_1}
\end{table}

\section{Decay time of overtones and ringdown of BBH mergers}
\label{sec:decaytime}
\begin{figure}[H]
  \begin{center}
    \includegraphics[keepaspectratio=true,height=55mm]{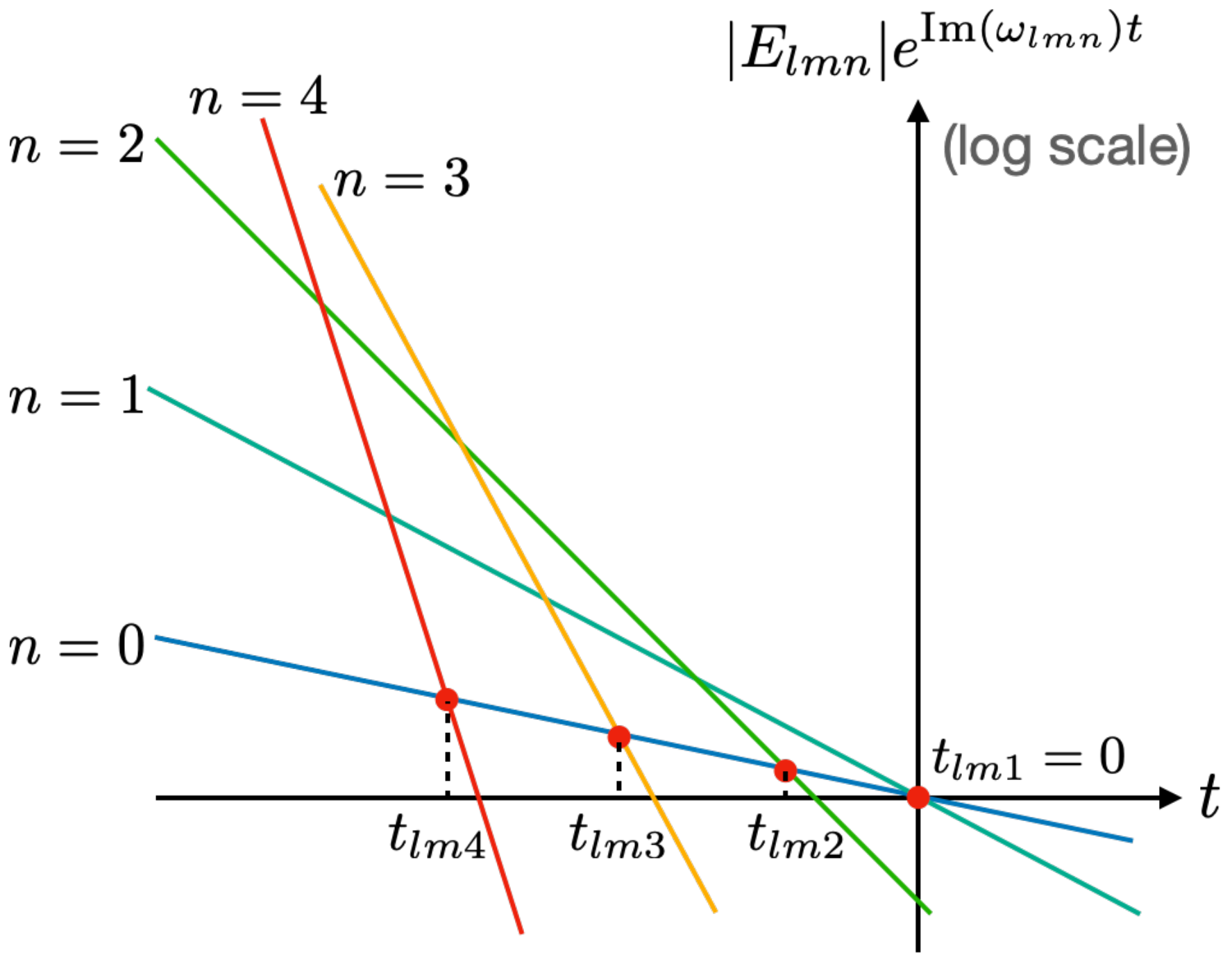}
  \end{center}
\caption{A schematic description of the definition of the decay time $t_{lmn}$ defined in (\ref{decay_time1}).
}
\label{decay_time_cartoon_fig}
\end{figure}

\begin{figure}[h]
  \begin{center}
    \includegraphics[keepaspectratio=true,height=50mm]{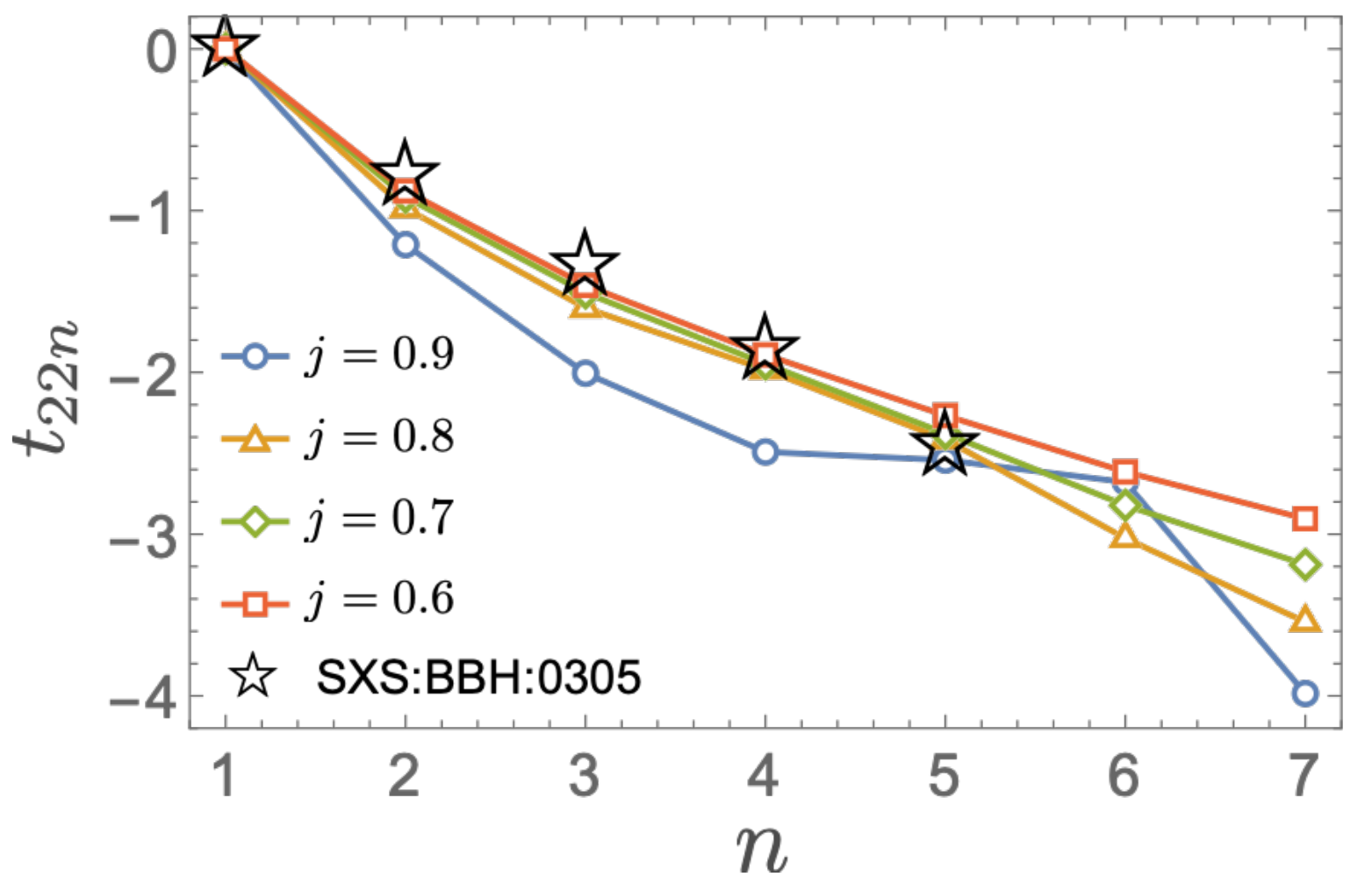}
  \end{center}
\caption{The decay time, $t_{22n}$, for $j=0.6$, $0.7$, $0.8$, and $0.9$. The star markers are the decay time estimated from the excitation coefficients extracted from the numerical relativity waveform SXS:BBH:0305 in Ref. \cite{Giesler:2019uxc}.}
\label{decay_time_fig}
\end{figure}

Although the QNEF and source factor are important to determine the amplitude of each QN mode, those factors may have the ambiguity of the ringdown start time, leading to an uncertainty of the factor $e^{-i\omega_{lmn} t_s}$. As such, we introduce a new quantity that is determined only by $E_{lmn}$ and $\omega_{lmn}$ and is independent of the ringdown start time. Since we know the absolute values of the QNEFs quantifying the typical amplitude of QN modes, one can estimate the time when the $n$-th overtone starts to be less dominant than the fundamental mode. That is independent of the ringdown start time and is important to properly determine the fit start time in the fitting data analysis. Let us introduce the following quantity, referred as {\it decay time}, for the $n$-th overtone
\begin{equation}
t_{lmn} \equiv \frac{\ln{|E_{lmn}/E_{lm0}|}}{{\rm Im}(\omega_{lmn}) - {\rm Im}(\omega_{lm0})} - \frac{\ln{|E_{lm1}/E_{lm0}|}}{{\rm Im}(\omega_{lm1}) - {\rm Im}(\omega_{lm0})}.
\label{decay_time1}
\end{equation}
For $t_{lmn} < t$, the $n$-th overtone of $(l,m)$ mode tends to be less important than the fundamental mode, and for $t > t_{lm1} = 0$, the fundamental mode dominates the signal (FIG. \ref{decay_time_cartoon_fig}). The decay time is defined only by the QN frequencies and QNEFs, and therefore, $t_{lmn}$ is also independent of the source of perturbation. In FIG. \ref{decay_time_fig}, the decay time, $t_{22n}$, with $n=0, 1, 2, \cdots, 7$ is shown for the various spin parameters. It is found that the decay time is insensitive to the spin parameters for $0.6 \lesssim j \lesssim 0.8$ that is the typical spin parameter range of remnant black holes of BBH mergers.
In Ref. \cite{Giesler:2019uxc}, the authors have performed the fit of QN modes to the numerical relativity waveform SXS:BBH:0305 \cite{Boyle:2019kee}, whose remnant black hole has $j=0.6921$. They extracted the excitation coefficients $C_{22n}$ up to $n=7$ by the fitting data analysis. We estimate the values of decay time, $\tilde{t}_{22n}$, by replacing $E_{22n}$ in (\ref{decay_time1}) with $C_{22n}$ obtained in \cite{Giesler:2019uxc}. Such a replacement is a good approximation when $|T_{22n}|/|T_{220}| \sim {\cal O}(1)$ that will be shown later in Table \ref{table_source_factor}. Then we confirm that $\tilde{t}_{22n}$ that is shown in FIG. \ref{decay_time_fig} with star markers is consistent with the decay time in (\ref{decay_time1}) as in Table \ref{table_excitation_time}.
\renewcommand{\arraystretch}{1.2}
\begin{table}[H]
\begin{center}
\begin{tabular}{c c c c c c c c} 
 \hline
 $n$ & 1 & 2 & 3 & 4 & 5  \\ [0.5ex] 
 \hline\hline
   $t_{22n}$ ($j=0.6921$) & 0& $-$0.908 & $-$1.50 & $-$1.94 & $-$2.37 \\ 
 \hline
 $\tilde{t}_{22n}$ (SXS:BBH:0305) &0& $-$0.792 & $-$1.35 & $-$1.88 & $-$2.47  \\
 \hline
\end{tabular}
\end{center}
\caption{The decay time, $t_{22n}$, for $j=0.6921$ and the decay time, $\tilde{t}_{22n}$, approximately evaluated by the excitation coefficients, $C_{22n}$, extracted from the numerical relativity waveform SXS:BBH:0305 ($j = 0.6921$) in Ref. \cite{Giesler:2019uxc}.}
\label{table_excitation_time}
\end{table}
\renewcommand{\arraystretch}{1}

The approximated decay time, $\tilde{t}_{lmn}$, computed by replacing the QNEF in (\ref{decay_time1}) with the excitation coefficient, has the log dependence of $T_{lmn}$, and therefore, such an approximation works only when
\begin{equation}
|T_{lmn}|/ |T_{lm0}| \sim {\cal O} (1).
\label{source_ratio}
\end{equation}
At least from our computation of $E_{22n} (j=0.6921)$ and from the fitting data analysis of SXS:BBH:0305 in \cite{Giesler:2019uxc}, one can check that the condition (\ref{source_ratio}) is satisfied while the $n$-dependence of $E_{22n}$ is significant as is shown in Table \ref{table_source_factor}, where $|T_{22n}|$ is computed as $|T_{22n}| = |C_{22n}|/|E_{22n}|$. In the future work, we will study the issue of to what extent the condition of (\ref{source_ratio}) holds for the ringdown signals sourced by BBH mergers. In Ref. \cite{Ma:2021znq}, it was found that the hierarchy and absolute values of the excitation coefficients of ringdown signals of the {\it superkick} configuration of BBH mergers (those with equal mass and anti-parallel spins) are insensitive to initial conditions. Their result is also well consistent with the behaviour of QNEFs and with that the source factors are insensitive to the overtone number (Table \ref{table_source_factor}).
\begin{table}[H]
\begin{center}
\begin{tabular}{c c c c c c c c c} 
 \hline
 $n$ &0 & 1 & 2 & 3 & 4 & 5 & 6 & 7 \\ [0.5ex] 
 \hline\hline
   $|E_{22n}|$ & 0.135& 0.546 & 1.26 & 2.21 & 3.13 & 3.31 & 2.69 & 1.98 \\ 
 \hline
  $|T_{22n}|$ & 7.21& 7.72 & 8.96 & 10.4 & 10.5 & 8.76 & 5.21 & 1.47 \\ 
 \hline
   $|E_{22n}|/|E_{220}|$ & 1& 4.06 & 9.37 & 16.4 & 23.3 & 24.6 & 20.0 & 14.7 \\ 
   \hline
   $|T_{22n}|/|T_{220}|$ & 1& 1.07 & 1.24 & 1.44 & 1.46 & 1.21 & 0.72 & 0.203 \\ 
 \hline
\end{tabular}
\end{center}
\caption{The absolute values of QNEFs for $j=0.6921$ and source factors estimated from the data of the excitation coefficients obtained in Ref. \cite{Giesler:2019uxc} as $T_{22n} = C_{22n}/E_{22n}$.}
\label{table_source_factor}
\end{table}
%%%%%%%%%%%%%%%%%%%%%%%%%
%%%%%%%%%%%%%%%%%%%%%%%%%
%%%%%%%%%%%%%%%%%%%%%%%%%
%%%%%%%%%%%%%%%%%%%%%%%%%

\section{Conclusion and discussion}
\label{sec:conclusion}
For the first time, we have computed the quasi-normal excitation factors (QNEF) that quantify the {\it ease of excitation} of QN modes of a Kerr black hole up to the 20th overtone for $0.3 \leq j \leq 0.99$. Then we have found that the first three highest QNEFs are $n=4$, $5$, and $6$ for the typical spin parameters of remnant black holes resulting from binary black hole (BBH) mergers. At $j=0.7$ which is close to the spin parameter of the remnant black hole of GW150914, we found that the first two highest values of $|E_{22n}|$ are at $n=4$ and $5$, which is consistent with the data analysis of GW150914-like numerical relativity waveform SXS:BBH:0305 in Ref. \cite{Giesler:2019uxc} and the recent fitting analysis that has investigated the universality of the importance of overtones \cite{Ma:2021znq}. Utilizing the result in Ref. \cite{Giesler:2019uxc}, we have found that the $n$-dependence of $T_{22n}$ in SXS:BBH:0305 is smaller than that of $E_{22n}$. Therefore, we conclude that the $n$-dependence of $C_{22n}$ is determined mostly by $E_{22n}$ at least in the waveform.
Also, we have shown that the overtone number for which the QNEF has the highest absolute value matches the overtone number at which the path of QN frequencies has a corner for $l=m=2$ (see FIG. \ref{f_corner}). We will investigate if this holds even for higher angular modes elsewhere. Indeed, the investigation of $E_{lmn}$ for higher angular modes are important to theoretically understand the significance of $l=m=3$ (c.f. \cite{Capano:2021etf}). In the latter part of Sec. \ref{sec:QNEF_result}, we have shown that the 5th QNEF $E_{225}$ is strongly suppressed than the other factors in the near-extremal situation as shown in FIG. \ref{QNEF_5th} and \ref{QN_mode_099}. This anomalous behaviour may correspond to the isolation of the 5th QN frequency in the near-extremal situation that has been found by Onozawa \cite{Onozawa:1996ux}. As far as I know, the physical reason of the mysterious behaviour of the 5th QN mode is still an open question.
In Sec. \ref{sec:plungeGW}, we have performed the numerical computation of the strain amplitude of GW signal induced by a particle plunging into a spinning black hole with corotating or counter rotating trajectories. Then we fit QN modes up to the 7th overtone to the waveforms, and we have found that the qualitative behaviour of the excitation coefficients, $C_{22n}$, agrees with that of the QNEFs. For example, we have confirmed that the first two highest values of the excitation coefficients are at $n=4$ and $5$ for $j=0.7$, and are at $n=5$ and $6$ for $j=0.9$. Also, we have confirmed that the excitation coefficient of the 5th overtone is strongly suppressed for the near-extremal case ($j=0.99$). All these results obtained by the extraction of the excitation coefficients from the GW signals are consistent with the direct computation of the QNEFs. Note that the extraction of the excitation coefficients in Sec. \ref{sec:plungeGW} have been performed in a manner totally independent of the computation of $E_{22n}$ in Sec. \ref{sec:QNEF_result}. Our results justify the truncation at $n=7$ commonly applied to the fit of QN modes to GW waveforms in the previous studies \cite{Giesler:2019uxc,Bhagwat:2019dtm,Dhani:2020nik,Finch:2021iip}. Also, it should be emphasized that depending on the initial data of BBH mergers, other overtones or fundamental mode could be more dominant than the 4th, 5th, or 6th overtones at the early ringdown although the source factor seems to be insensitive to most of the initial conditions.

In Sec. \ref{sec:decaytime}, we have introduced the decay time of QN mode that is determined only by the QN frequencies and QNEFs. It is useful to predict the time when the $n$-th QN mode tends to be suppressed compared to the fundamental QN mode and to determine the fit start time of the fitting data analysis.
Our computation is based on the fact that the solution of the Teukolsky equation for the KdS spacetime is represented by the general Heun function, that was discovered by Suzuki, Takasugi, and Umetsu \cite{Suzuki:1998vy}. The Heun function is available in Mathematica 12.1 or later version, and it makes possible to compute QN frequencies of the KdS spacetime with high accuracy. In the limit of zero cosmological constant, one can obtain the QN frequencies of the Kerr spacetime as was performed in \cite{Hatsuda:2020sbn}. We have extended this technique to the computation of QNEFs and overtones as described in Sec. \ref{sec:methodology}.

The excitation of overtones should be quantified by the excitation coefficients $C_{lmn}$ since $C_{lmn} S_{lmn}$ gives the amplitude of each QN mode. However, it should be emphasized that each QN mode can be characterized by its ease of excitation, and it is determined only by the intrinsic nature of the black hole. In other words, each QN mode is characterized not only by its unique frequency $\text{Re}(\omega_{lmn})$ and damping rate $\text{Im}(\omega_{lmn})$ but also by the QNEF $E_{lmn}$ independent of the source of perturbation. What has been done in this paper is quantifying the ease of excitation of QN modes of Kerr black holes up to higher overtones $n\leq 20$. On the other hand, a {\it direct} computation of the source factors would be significantly challenging especially for non-linearly distorted black holes. Nevertheless, combining our result of $E_{22n}$ and the excitation coefficient extracted by the fitting data analysis, one can obtain the source factors even from GW signals of BBH mergers as we performed in Sec. \ref{sec:decaytime}. This procedure to extract the source factors may contribute to the progress in the modelling of GW waveforms for linearly perturbed black holes. However, we should note that the excitation of QN modes at or before the strain peak is still controversial especially for comparable mass-ratio BBH mergers since it may involve the highly non-linear regime. As an independent check if a remnant black hole settle to perturbative state as early as the strain peak, one can see the time evolution of the precession of the remnant as was performed in \cite{Hamilton:2021pkf}. If the remnant can be described by a perturbed Kerr black hole at or before the strain peak, its dynamical precession would be suppressed at the moment.

Recently, the importance of the mirror overtones \cite{Dhani:2020nik,Dhani:2021vac} and the excitation of $l=m=3$ harmonics \cite{Capano:2021etf} have been investigated. As an interesting extension of our work, one can discuss the importance of mirror and higher angular modes by computing QNEFs up to higher overtones (QNEFs up to the 3rd overtone are available in \cite{Berti:cite,grit:cite}). We leave the analysis of the mirror modes or higher angular modes based on the QNEF for future studies.

\begin{acknowledgements}
The author thanks Niayesh Afshordi for fruitful discussions and helpful comments. The author is also grateful to Sizheng Ma and Lionel London for sharing their important works and giving some interesting comments that stimulate the future research on the ringdown modeling. The author is supported by the Special Postdoctoral Researcher (SPDR) Program at RIKEN, FY2021 Incentive Research Project at RIKEN, and Grant-in-Aid for Scientific Research (KAKENHI) project for FY 2021 (21K20371).
\end{acknowledgements}

\appendix
\section{SWSH factor $S_{lmn}$ and its dependence on the overtone number $n$}
\label{app:SWSHF}
The amplitude of each QN mode is given by the product of the QNEF, $E_{lmn}$, source factor, $T_{lmn}$, and the SWSH factor $S_{lmn}$. The two factors, $E_{lmn}$ and $S_{lmn}$, are independent of the source of perturbation. Nevetheless, the $n$-dependence of the amplitude of QN mode is determined mostly by $E_{lmn}$ because the $n$-dependence of the SWSH factor is smaller than that of the QNEF as is shown below. In FIG. \ref{SWSH_zero_plot}, the SWSH factor for the fundamental QN mode, $S_{220}$, is shown for the various spin parameters. One can see that the spin dependence of $S_{220}$ is small. Also, the SWSH factor has the maximum value at $\theta = 0$ and is suppressed at $\theta = \pi$. On the other hand, FIG. \ref{SWSH_ratio_plot} shows the absolute values of the ratio $S_{22n} (\theta)/ S_{220} (\theta)$ from which one can read that it is of the order of unity in the whole range of $\theta$ and for $n=0, 1, 2, \cdots, 10$ while the ratio of the QNEFs, $E_{22n}/E_{220}$, is at most of the order of $10^2$ (see FIG. \ref{QNEF_5th}). Therefore, we conclude that the dependence of the ease of excitation of QN mode on the overtone number, $n$, is determined mostly by the QNEF. Of course, taking into account the SWSH factor is necessary to estimate the amplitude of QN modes. Here we utilized the Mathematica notebook available in \cite{Berti:cite,grit:cite} to compute the SWSH factor although the values of QN frequencies are obtained by the independent methodology using the general Heun function.
%%%%%%%%%%%%%%%%%%%%%%%%%
\begin{figure}[h]
  \begin{center}
    \includegraphics[keepaspectratio=true,height=100mm]{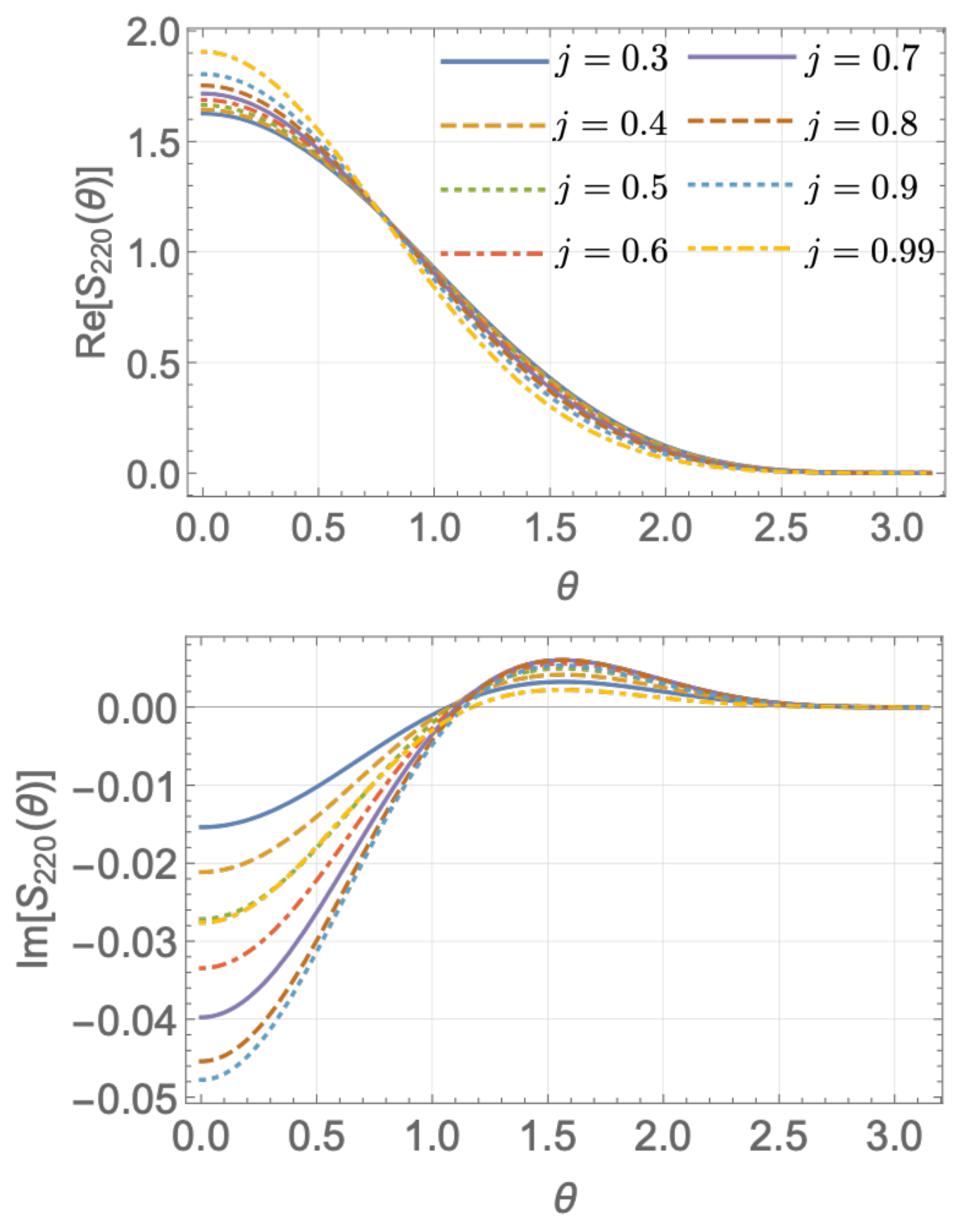}
  \end{center}
\caption{The SWSH factor for the fundamental QN mode, $S_{220} (\theta)$, is shown for $j=0.3$, $0.4$, $0.5$, $0.6$, $0.7$, $0.8$, $0.9$, and $0.99$.}
\label{SWSH_zero_plot}
\end{figure}
%%%%%%%%%%%%%%%%%%%%%%%%%
%%%%%%%%%%%%%%%%%%%%%%%%%
\begin{figure}[h]
  \begin{center}
    \includegraphics[keepaspectratio=true,height=110mm]{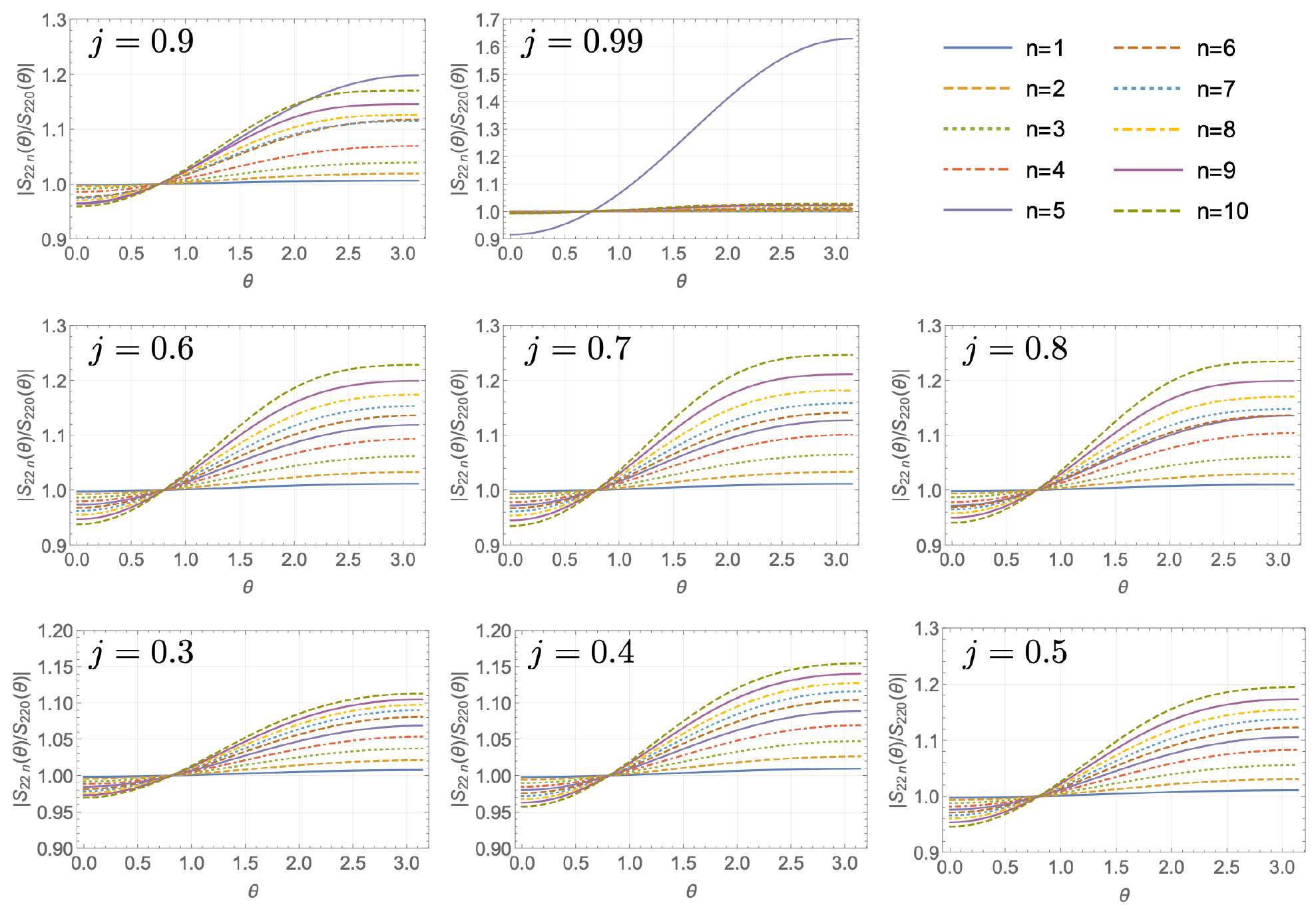}
  \end{center}
\caption{The absolute values of the ratio of the $n$-th SWSH factor to the zeroth one, $|S_{22n} (\theta)/ S_{220} (\theta)|$, is shown for the same spin parameters in FIG.\ref{SWSH_zero_plot}.}
\label{SWSH_ratio_plot}
\end{figure}
%%%%%%%%%%%%%%%%%%%%%%%%%

\section{Computation of QN frequencies}
\label{app:QN frequencies}
\begin{figure}[h]
  \begin{center}
    \includegraphics[keepaspectratio=true,height=70mm]{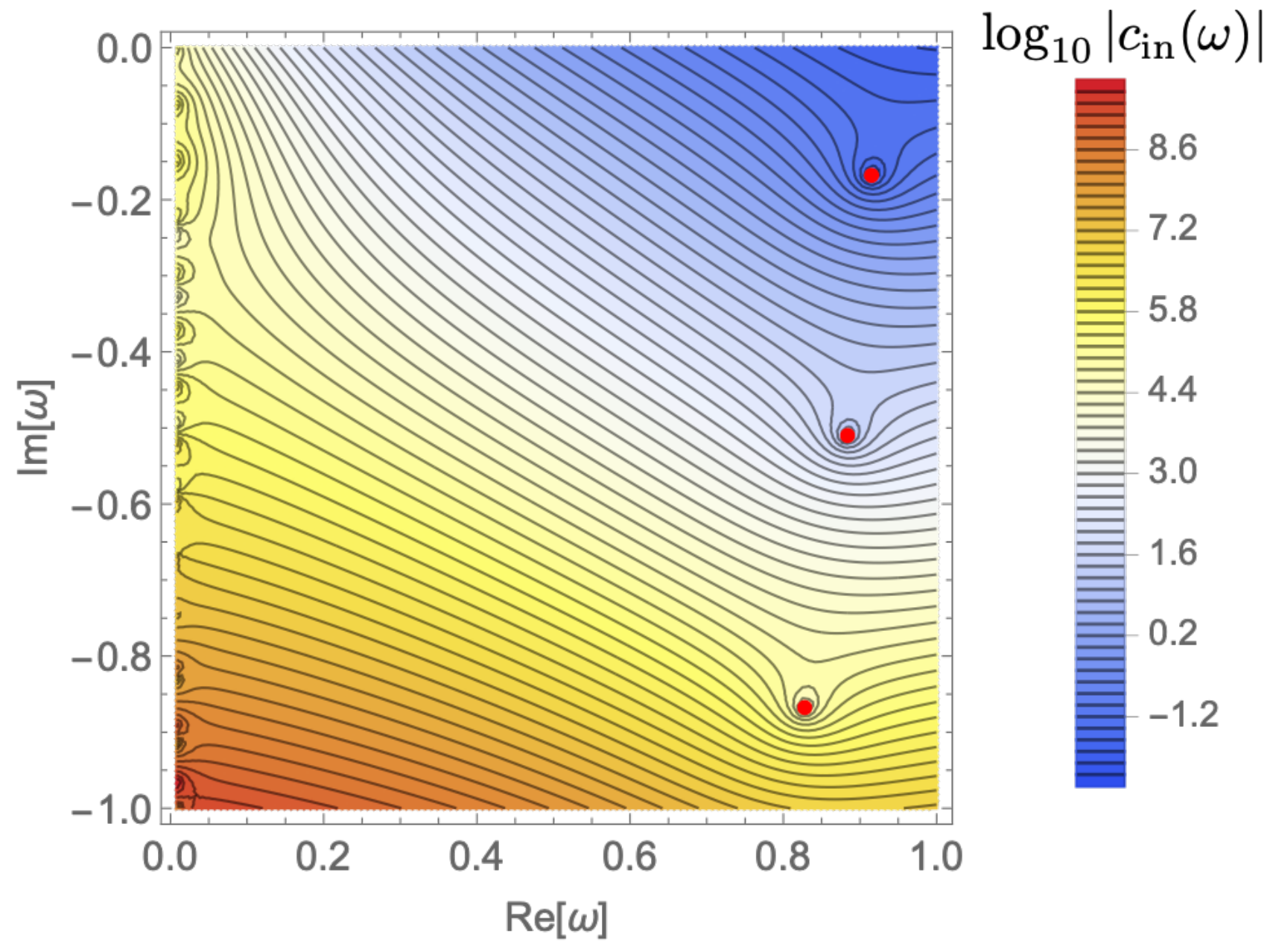}
  \end{center}
\caption{Contour plot showing the values of $\log_{10} |c_{\inn} (\omega)|$ in the complex frequency plane for $M=0.5$, $j=0.25$, $l=m=2$, and $3H^2 = 0.02$. The first three QN frequencies for which $c_{\inn} = 0$ are indicated by the red points.}
\label{KDSQNM_plot}
\end{figure}

It is known that the solution of the Teukolsky equation in the KdS spacetime can be represented by the Heun function \cite{Suzuki:1998vy}. This allows us to precisely compute the QN modes of Kerr spacetime by extrapolating the QN modes of KdS spacetime to $H = 0$ \cite{Hatsuda:2020sbn}, where $H$ is the Hubble parameter. Here we review the methodology to compute the QN modes in the KdS spacetime. Then in the next section, we will present the details of how to compute the QNEF of the Kerr spacetime. The angular and radial Teukolsky equations for a spin-$s$ field are
\begin{align}
\left[ \frac{d}{du} \Delta_u \frac{d}{du} - \frac{1}{\Delta_u} \left(  V + \frac{s}{2} \Delta_u' \right)^2 + 2s V' - X_s \right] {}_{s} S_{\omega l m} (u) &= 0, \label{angular_T}\\
\left[ \Delta_r^{-s} \frac{d}{dr} \Delta_r^{s+1} \frac{d}{dr} + \frac{1}{\Delta_r} \left( W^2 -isW \Delta_r' \right) + 2is W' - Y_s \right]{}_{s} R_{\omega l m} (r) &= 0, \label{radial_T}
\end{align}
and
\begin{align}
V(u) &\equiv \Xi \left[ a\omega (1-u^2) - m \right],\\
W(r) &\equiv \Xi \left[ \omega (r^2+a^2) -am \right],\\
X_s(u) &\equiv 2 (2s^2 + 1) \alpha^2 u^2 -  \lambda,\\
Y_s(r) &\equiv 2 H^2 (s+1) (2s+1) r^2 + \lambda -s (1-\alpha^2),\\
\Delta_r \equiv (r^2+a^2) (1-H^2 r^2) -2Mr, \\
\Delta_u &\equiv (1-u^2) (1+\alpha^2 u^2),\\
u &\equiv \cos \theta,\\
\alpha &\equiv a^2 H^2,
\end{align}
where $\lambda = \lambda(\omega, \alpha)$ is the separation constant. Let us consider the following transformations of ${}_s S_{\omega lm}$ and ${}_s R_{\omega lm}$
\begin{align}
{}_s S_{\omega lm} (u) &= f^{A_1} (f-1)^{A_2} (f-f_a)^{A_3} (f-f_{\infty}) y_a(f),\\
f &= f (u) \equiv \frac{(1-i/\sqrt{\alpha}) (u+1)}{2 (u-i/\sqrt{\alpha})}, \
f_a \equiv - \frac{(1-i/\sqrt{\alpha})^2}{4i/\sqrt{\alpha}}, \
f_{\infty} \equiv \frac{1-i / \sqrt{\alpha}}{2},\\
{}_s R_{\omega lm} (r) &= g^{B_1} (g-1)^{B_2} (g-g_r)^{B_3} (g-g_{\infty})^{2s+1} y_{r} (g),\\
g & = g(r) \equiv \frac{(r_{c+}-r_-) (r-r_+)}{(r_{c+} -r_+) (r-r_-)}, \
g_r \equiv \frac{(r_{c+}-r_-) (r_{c-}-r_+)}{(r_{c+} -r_+) (r_{c-}-r_-)}, \
g_{\infty} \equiv \frac{r_{c+}-r_-}{r_{c+} -r_+},
\end{align}
where $r_{c\pm}$ and $r_{\pm}$ are the roots of $\Delta (r) = 0$ with $r_{c-} < r_- < r_+ < r_{c+}$ and
\begin{align}
&A_1 \equiv \frac{m-s}{2}, \ A_2 \equiv \frac{m+s}{2}, \ A_3 \equiv \frac{i}{2} \left( \frac{1+\alpha}{\sqrt{\alpha}} c -m\sqrt{\alpha} -is \right), \ c\equiv a \omega\\
&B_1 \equiv \frac{i (1+\alpha) K(r_+)}{\Delta' (r_+)}, \ B_2 \equiv \frac{i (1+\alpha) K(r_{c+})}{\Delta' (r_{c+})}, \ B_3 \equiv \frac{i (1+\alpha) K(r_{c-})}{\Delta' (r_{c-})}.
\end{align}
The black hole horizon and cosmological horizon are at $r=r_+$ ($g=0$) and at $r=r_{c+}$ ($g=1$), respectively. Performing the transformations, the angular and radial equations (\ref{angular_T}) and (\ref{radial_T}) reduce to
\begin{align}
&\frac{d^2 y_a (f)}{df ^2} + \left( \frac{2A_1 + 1}{f} + \frac{2A_2+1}{f-1} + \frac{2A_3 + 1}{f-f_a} \right) \frac{d y_a (f)}{df} + \frac{\rho_{a+} \rho_{a -} f - U_a}{f (f-1) (f-f_a)} y_a (f) = 0,\label{angular_heun_eq}\\
&\frac{d^2 y_r (g)}{dg ^2} + \left( \frac{2B_1 +s + 1}{g} + \frac{2B_2+s+1}{g-1} + \frac{2B_3 +s+ 1}{g-g_a} \right) \frac{d y_r (g)}{dg} + \frac{\rho_{r+} \rho_{r -} g - U_r}{g (g-1) (g-g_a)} y_r (g) = 0,
\label{radial_heun_eq}
\end{align}
where
\begin{align}
&\rho_{a+} \equiv 1, \ \rho_{a-} \equiv 1-s -im \sqrt{\alpha} + ic \left( \sqrt{\alpha} + \frac{1}{\sqrt{\alpha}} \right),\\
&U_{a} \equiv \frac{i \lambda}{4 \sqrt{\alpha}} + \frac{1}{2} + A_1 + \left( m+\frac{1}{2} \right) (A_3 - A_3^{\ast}),\\
&\rho_{r+} \equiv 2s+1, \ \rho_{r-} \equiv s+1 - \frac{2 i (1+\alpha) K(r_-)}{\Delta'(r_-)},\\
\begin{split}
&U_r \equiv -\frac{(1+s) (1+2s)r_{c-}}{r_- - r_{c-}} - \frac{\lambda -2s (1-\alpha) +H^2 (1+s) (1+2s) r_+ (r_+ + r_{c+})}{H^2 (r_- - r_{c-}) (r_+ - r_{c+})}\\
&~~~~~~~ + \frac{2i (1+2s) (1+\alpha) (r_+ r_- \omega+a^2 \omega - am)}{H^2 (r_- - r_{c-}) (r_--r_+) (r_+ - r_{c+})}.
\end{split}
\end{align}
Both the equations, (\ref{angular_heun_eq}) and (\ref{radial_heun_eq}), are the Heun's differential equation:
\begin{align}
&y''(x) + \left( \frac{\zeta_1}{x} + \frac{\zeta_2}{x-1} + \frac{\zeta_3}{x-x_a} \right) y'(x) + \frac{\rho_+ \rho_- x - U}{x(x-1) (x-x_a)} y(x) = 0,
\label{heun_eq_example}\\
\text{with} \ &\zeta_1 + \zeta_2 + \zeta_3 = \rho_+ + \rho_- + 1.
\end{align}
Here we are interested in the SWSH function, ${}_s S_{\omega lm} (\theta)$, and the Teukolsky variable, ${}_s R_{\omega lm} (r)$, which satisfy the proper boundary conditions at $\theta = \pi$, $\theta = 0$ ($f=0$, $f=1$) and at $r= r_+$, $r= r_{c+}$ ($g=0$, $g=1$), respectively. In the following, as such, let us consider the solution of the Heun's differential equation (\ref{heun_eq_example}) which is convergent near $x=0$ or $x=1$. For each point, it has the form of
\begin{align}
y(x) =
\begin{cases}
c_{01} y_{01} (x) + c_{02} y_{02} (x) & \text{for} \ x \sim 0,\\
c_{11} y_{11} (x) + c_{12} y_{12} (x) & \text{for} \ x \sim 1,
\end{cases}
\end{align}
with
\begin{align}
y_{01} (x) &\equiv H\ell (x_a, U; \rho_+, \rho_-, \zeta_1, \zeta_2, x),\label{y01}\\
y_{02} (x) &\equiv x^{1-\zeta_1} H\ell (x_a, (x_a \zeta_2+\zeta_3) (1-\zeta_1)+U; \rho_+ +1- \zeta_1, \rho_- + 1- \zeta_1, 2-\zeta_1, \zeta_2, x),\label{y02}\\
y_{11} (x) &\equiv H\ell (1-x_a, \rho_+ \rho_- - U; \rho_+, \rho_-, \zeta_2, \zeta_1, 1-x),\label{y11}\\
\begin{split}
y_{12} (x) &\equiv (1-x)^{1-\zeta_2}\\
&~~~\times H\ell (1-x_a, [(1-x_a) \zeta_1 + \zeta_3] (1-\zeta_2) + \rho_+ \rho_- - U; \rho_+ +1- \zeta_2, \rho_- + 1- \zeta_2, 2-\zeta_2, \zeta_1, 1-x),
\end{split}
\label{y12}
\end{align}
where $H\ell$ is the symbol of the general Heun's function, and $c_{01}$, $c_{02}$, $c_{11}$, and $c_{12}$ are constants. The solution of (\ref{angular_heun_eq}) near $f=0$ or $f=1$ is obtained by the following identification in (\ref{y01})-(\ref{y12}):
\begin{align}
& x\to f,\\
&\zeta_i \to 2A_i + 1 \ \text{with} \ i=1,2,3,\\
&\rho_{\pm} \to \rho_{a\pm}, \ U \to U_a, \ x_a \to f_a.
\end{align}
Then the boundary condition of the SWSH function at $f=0$ ($u=-1$) is
\begin{align}
{}_s S_{\omega lm} \propto
\begin{cases}
f^{A_1} (f-1)^{A_2} (f-f_a)^{A_3} (f-f_{\infty}) y_{01} (f) \sim (1+u)^{(m-s)/2}, \ &\text{for} \ m-s \geq 0,\\
f^{A_1} (f-1)^{A_2} (f-f_a)^{A_3} (f-f_{\infty}) y_{02} (f) \sim (1+u)^{-(m-s)/2}, \ &\text{for} \ m-s \leq 0,
\end{cases}
\end{align}
and at $f=1$ ($u=1$)
\begin{align}
{}_s S_{\omega lm} \propto
\begin{cases}
f^{A_1} (f-1)^{A_2} (f-f_a)^{A_3} (f-f_{\infty}) y_{11} (f) \sim (1-u)^{-(m+s)/2}, \ &\text{for} \ m+s \leq 0,\\
f^{A_1} (f-1)^{A_2} (f-f_a)^{A_3} (f-f_{\infty}) y_{12} (f) \sim (1-u)^{(m+s)/2}, \ &\text{for} \ m+s \geq 0.
\end{cases}
\end{align}
The eigenvalue ${}_s \lambda_{lm}$ is obtained by requiring the regularity of the SWSH function, ${}_sS_{lm}$, at $f=0$ and $f=1$. For example, for the angular mode $l=m=2$ of gravitational field $s=-2$, the eigenvalue, $\lambda = {}_{-2} \lambda_{22}$, is obtained by solving
\begin{equation}
{\cal W}[y_{01} (f), y_{11} (f)] = y_{01} \frac{d y_{11}}{df} - \frac{d y_{01}}{df} y_{11} =0,
\end{equation}
with respect to $\lambda$. Note that the eigenvalue ${}_s \lambda_{lm}$ reduces to $l (l+1) -s (s-1)$ for $c \to 0$ and $\alpha \to 0$. The solution of radial Teukolsky equation, ${}_s R_{\omega lm}$, is also given by the linear combination of (\ref{y01}) and (\ref{y02}) at $g=0$ and given by that of (\ref{y11}) and (\ref{y12}) at $g=1$ with the identification of
\begin{align}
&x\to g,\\
&\zeta_i \to 2B_i +s+ 1 \ \text{with} \ i=1,2,3,\\
&\rho_{\pm} \to \rho_{r\pm}, \ U \to U_r, \ x_a \to g_a,
\end{align}
in (\ref{y01})-(\ref{y12}). Then the ingoing and outgoing modes near the black hole horizon ($g = 0$) are
\begin{align}
R_{\inn}^{\bh} (g) &= g^{B_1} (g-1)^{B_2} (g-g_r)^{B_3} (g-g_{\infty}) y_{02} (g) \propto (r-r_+)^{-s} \exp{\left[ -i (1+\alpha) (\omega-m\Omega_{\rm H}) r^{\ast} \right]},\\
R_{\out}^{\bh} (g) &= g^{B_1} (g-1)^{B_2} (g-g_r)^{B_3} (g-g_{\infty}) y_{01} (g) \propto \exp{\left[ +i (1+\alpha) (\omega-m\Omega_{\rm H}) r^{\ast} \right]}.
\end{align}
and the ingoing and outgoing ones near the cosmological horizon ($g=1$) are
\begin{align}
R_{\inn}^{\co} (g) &= g^{B_1} (g-1)^{B_2} (g-g_r)^{B_3} (g-g_{\infty}) y_{12} (g) \propto (r_{c+} -r)^{-s} \exp{\left[ -i (1+\alpha) (\omega-m \Omega_{\rm C}) r^{\ast} \right]},\\
R_{\out}^{\co} (g) &= g^{B_1} (g-1)^{B_2} (g-g_r)^{B_3} (g-g_{\infty}) y_{11} (g) \propto \exp{\left[ +i (1+\alpha) (\omega-m \Omega_{\rm C}) r^{\ast} \right]},
\end{align}
where $dr^{\ast}/dr \equiv (r^2+a^2)/ \Delta_r$, $\Omega_{\rm H} \equiv a/(r_+^2+a^2)$, and $\Omega_{\rm C} \equiv a/(r_{c+}^2+a^2)$. The ingoing modes near $g=0$, $R_{\inn}^{\bh}$, is given by the linear combination of $R_{\inn}^{\co}$ and $R_{\out}^{\co}$ near $g=1$
\begin{equation}
y_{02} (g) = c_{\out} (\omega) y_{11} (g) + c_{\inn} (\omega) y_{12} (g).
\end{equation}
Requiring $c_{\inn} (\omega) = 0$ for a specific value of complex frequency while imposing the regularity of the SWSH function, one obtains QN frequencies of the KdS spacetime.
FIG. \ref{KDSQNM_plot} shows the contour plot of $c_{\inn} (\omega)$ in the complex frequency plane and the zero points (indicated by red points in FIG. \ref{KDSQNM_plot}) corresponds to QN modes.
%%%%%%%%%%%%%%%%%%%%%%%%%
%%%%%%%%%%%%%%%%%%%%%%%%%

\section{Computation of QNEFs}
\label{app:QNEFs}

Let us consider a purely ingoing radial solution at the black hole horizon, $R_{\inn}^{\bh} (g)$, which is the superposition of ingoing and outgoing modes near the cosmological horizon
\begin{equation}
R_{\inn}^{\bh} (g) = g^{B1} (g-1)^{B2} (g-g_r)^{B3} (g-g_{\infty})^{2s+1} (c_{\out} y_{11} (g) + c_{\inn} y_{12} (g)).
\end{equation}
Its asymptotic behavior at the cosmological horizon ($g \to 1$) is
\begin{align}\displaystyle
\begin{split}
\lim_{g\to 1} R_{\inn}^{\bh} (g) &=c_{\out} (-1)^{B2} (1-g_r)^{B3} (1-g_{\infty})^{2s+1} (1-g)^{B_2} + c_{\inn} (-1)^{B2} (1-g_r)^{B3} (1-g_{\infty})^{2s+1} (1-g)^{-B_2-s},\\
&=c_{\out} {\cal Q}_{\out} \left(\frac{r_{c+}-r}{r_{c+} -r_{c-}} \right)^{B_2} + c_{\inn} {\cal Q}_{\inn} \left(\frac{r_{c+}-r}{r_{c+} -r_{c-}} \right)^{-s-B_2},
\end{split}\label{KDS_asymp_inf}\\
\text{with} \ &{\cal Q}_{\out}\equiv (-1)^{B2} (1-g_r)^{B3} (1-g_{\infty})^{2s+1} \left( \frac{(r_+-r_-)(r_{c+}-r_{c-})}{(r_{c+}-r_-) (r_{c+}-r_-)} \right)^{B_2},\\
&{\cal Q}_{\inn}\equiv (-1)^{B2} (1-g_r)^{B3} (1-g_{\infty})^{2s+1} \left( \frac{(r_+-r_-)(r_{c+}-r_{c-})}{(r_{c+}-r_-) (r_{c+}-r_-)} \right)^{-s-B_2}.
\end{align}
This is the asymptotic behavior at the cosmological horizon, but we are interested in the asymptotic form in the intermediate region $r_+ \ll r \ll r_{c+}$ with the flat limit of $r_{c+} \to \infty$. As such, we have to relate the asymptotic form at $r \to r_{c+}$ and that in $r_+ \ll r \ll r_{c+}$. To this end, let us consider perturbations on a pure de Sitter spacetime with a cosmological constant, $3H^2$, to read the relation between ingoing/outgoing modes at $r \simeq 1/H$ and the modes in $r \ll 1/H$.
The radial perturbation, $R_{\rm ds}$, is governed by the following equation \cite{Suzuki:1995nh}:
\begin{align}
\begin{split}
&z^2 (1-z^2)^2 \frac{d^2 R_{\rm ds}}{dz^2}  - [2 (-s+1)z^3 -2 (s+1)(1-z^2)z] (1-z^2) \frac{d R_{\rm ds}}{dz}\\
&- \left\{ (1-z^2) [(l-s)(l+s+1)+2 z^2]
-(z\omega/H)^2-2is z\omega /H \right\} R_{\rm ds} = 0,
\label{ds_pert_eq}
\end{split}
\end{align}
where $z \equiv Hr$. Implementing the following transformation,
\begin{align}
&z \to \xi \equiv \frac{1-z}{1+z},\\
&R_{\rm ds} \to f_{\rm ds} \equiv \xi^{p+i\omega/(2H)} (1-\xi)^{s-l} (1+\xi)^{-2s-1} R_{\rm ds},
\end{align}
the wave equation (\ref{ds_pert_eq}) reduces to the hypergeometric equation
\begin{equation}
\xi(1-\xi) \frac{d^2 f_{\rm ds}}{d\xi^2} + [1+s-i\omega/H -(2l+3+s-i\omega/H) \xi] \frac{d f_{\rm ds}}{d\xi} -(l+s+1) (l+1-i\omega/H) f_{\rm ds} = 0,
\end{equation}
and we have
\begin{align}
R_{\rm ds} =
\begin{cases}
R_{\rm ds}^{(\out)} \equiv \xi^{-s-i\omega/(2H)} (1-\xi)^{l-s} (1+\xi)^{2s+1} {}_2F_1 [1+l+s,1+l-i\omega/H,1+s-i\omega/H,\xi] & \text{(outgoing)},\\
R_{\rm ds}^{(\inn)} \equiv \xi^{i\omega/(2H)} (1-\xi)^{l-s} (1+\xi)^{2s+1} {}_2F_1 [1+l-s,1+l+i\omega/H,1-s+i\omega/H,\xi] & \text{(ingoing)}.
\end{cases}
\label{ds_pert_asym_sol}
\end{align}
In the limit of $H M \to 0$, the typical frequencies of black hole, $\omega \sim 1/M$, are much higher than $H$. Therefore, we can take the limit of $\omega/H \to \infty$ in (\ref{ds_pert_asym_sol}) and obtain the following behavior of $R_{\rm ds}$
\begin{align}
R_{\rm ds}^{(\out)} \simeq
\begin{cases}
\displaystyle
(2H^{-1})^{s+i\omega/(2H)}(H^{-1}-r)^{-s-i\omega/(2H)} & (r\to H^{-1}),\\
\displaystyle
\frac{H^{-2s-1}}{r^{2s+1}} e^{i\omega r} & (r \ll H^{-1}),
\end{cases}
\end{align}

\begin{align}
R_{\rm ds}^{(\inn)} \simeq
\begin{cases}
\displaystyle
(2H^{-1})^{-i\omega/(2H)}(H^{-1}-r)^{i\omega/(2H)} & (r\to H^{-1}),\\
\displaystyle
\frac{2^{2s}H^{-1}}{r} e^{-i\omega r} & (r \ll H^{-1}),
\end{cases}
\end{align}
where we use ${}_2F_1 [1+l+s,-i\omega/H,-i\omega/H, \xi] = (1-\xi)^{-1-l-s}$ and ${}_2F_1 [1+l-s,i\omega/H,i\omega/H, \xi] = (1-\xi)^{-1-l+s}$.
Finally, taking the limit of $HM \to 0$ in (\ref{KDS_asymp_inf}), we have
\begin{equation}\displaystyle
\lim_{MH \to 0} R_{\inn}^{(\text{BH})} (g) =
c_{\inn} {\cal Q}_{\inn} R^{(\inn)}_{\rm ds} + c_{\out} {\cal Q}_{\out} R^{(\out)}_{\rm ds},
\end{equation}
and at the intermediate region $r_+ \ll r \ll r_{c+}$,
\begin{equation}
\lim_{MH \to 0} R_{\inn}^{(\text{BH})} (g) =
\frac{A^{(\inn)}_{lm}}{r} e^{-i \omega r} + \frac{A^{(\out)}_{lm}}{r^{2s+1}} e^{i \omega r},
\end{equation}
with
\begin{align}
\begin{split}
A^{(\inn)}_{lm} = c_{\inn} {\cal Q}_{\inn} 2^{2s} H^{-1},\\
A^{(\out)}_{lm} = c_{\out} {\cal Q}_{\out} H^{-2s -1}.
\end{split}
\label{in_out_coefficients}
\end{align}
One can compute the QNEFs of the Kerr spacetime by substituting the coefficients (\ref{in_out_coefficients}) into (\ref{EF_formula}).

\section{QN freuqnecies of the Kerr spacetime up to the 20th overtones}
\label{app:QNtable}
Here we show the QN frequencies of the Kerr spacetime computed by the methodology described in Sec. \ref{sec:methodology}. Our result is in agreement with the catalog of QN modes provided in \cite{Berti:cite,grit:cite}, where the data of QN frequencies for $n = 0$, $1$, $2$, and $3$ are available.

\begin{table}
\begin{center}
\begin{tabular}{c|ll|ll|ll|ll}
\firsthline
 &\multicolumn{2}{c|}{$j=0.7$} &\multicolumn{2}{c|}{$j=0.8$}&\multicolumn{2}{c|}{$j=0.9$}&\multicolumn{2}{c}{$j=0.99$} \\
\cline{2-9}
~$n$~    & Re$(\omega_{22n})$ & Im$(\omega_{22n})$ & Re$(\omega_{22n})$ & Im$(\omega_{22n})$ & Re$(\omega_{22n})$ & Im$(\omega_{22n})$ & Re$(\omega_{22n})$ & Im$(\omega_{22n})$ \\
\hline
$0$      & 1.065200487 & $-$0.1615857463 & 1.172033950 & $-$0.1512591047 & 1.343228544 & $-$0.1297384718 &  1.741785317 & $-$0.05878084844 \\
$1$      & 1.042321531 & $-$0.4884766316 & 1.155844794 & $-$0.4562978801 & 1.335315102 & $-$0.3905041341 & 1.741290328 & $-$0.1763507666 \\
$2$      & 0.9998125157 & $-$0.8245230723 & 1.124479630 & $-$0.7677904212 & 1.319653364 & $-$0.6550367507 & 1.740335349 & $-$0.2939311351 \\
$3$      & 0.9426725977 & $-$1.168601432 & 1.077911959 & $-$1.085776247 & 1.295738067 & $-$0.9257284931 & 1.739029926 & $-$0.4115377810 \\
$4$      & 0.8807699813 & $-$1.507599307 & 1.012525971 & $-$1.395924771 & 1.259672117 & $-$1.206590109 & 1.737532974 & $-$0.5292613212 \\
$5$      & 0.8469007434 & $-$1.835899064 & 0.9725668689 & $-$1.661605321 & 1.073833442 & $-$1.497446926 & 1.012863751 & $-$1.422766126 \\
$6$      & 0.8477071310 & $-$2.190815052 & 0.9983298183 & $-$1.966167230 & 1.204454964 & $-$1.540673954 & 1.735986660 & $-$0.6472760312 \\
$7$      & 0.8510746470 & $-$2.574238877 & 1.013303241 & $-$2.313415121 & 1.237597676 & $-$1.866356633 & 1.734473025 & $-$0.7657964462 \\
$8$      & 0.8505188425 & $-$2.970291020 & 1.017962117 & $-$2.671328320 & 1.250689150 & $-$2.163941149 & 1.733018798 & $-$0.8850187394 \\
$9$      & 0.8477216126 & $-$3.372527324 & 1.018408101 & $-$3.034004217 & 1.256539957 & $-$2.460017072 & 1.731635959 & $-$1.005079631 \\
$10$      & 0.8439056065 & $-$3.778516304 & 1.016956808 & $-$3.399745589 & 1.259128247 & $-$2.756833614 & 1.730365773 & $-$1.126038608 \\
$11$      & 0.8396771422 & $-$4.187119221 & 1.014596352 & $-$3.767742860 & 1.259980000 & $-$3.054746466 & 1.729298064 & $-$1.247868625 \\
$12$      & 0.8353376606 & $-$4.597709004 & 1.011810702 & $-$4.137513162 & 1.259831459 & $-$3.353733666 & 1.728551020 & $-$1.370447702 \\
$13$      & 0.8310514738 & $-$5.009905414 & 1.008872099 & $-$4.508739782 & 1.259091775 & $-$3.653688856 & 1.728216414 & $-$1.493568487 \\
$14$      & 0.8269266230 & $-$5.423466081 & 1.005962050 & $-$4.881204944 & 1.258014458 & $-$3.954497139 & 1.728305804 & $-$1.616987060 \\
$15$      & 0.8230573992 & $-$5.838234064 & 1.003230118 & $-$5.254752607 & 1.256773390 & $-$4.256054607 & 1.728743185 & $-$1.740493714 \\
$16$      & 0.8195528629 & $-$6.254108284 & 1.000829413 & $-$5.629260719 & 1.255501216 & $-$4.558270079 & 1.729406439 & $-$1.863957491 \\
$17$      & 0.8165654448 & $-$6.671020907 & 0.9989443119 & $-$6.004609649 & 1.254310731 & $-$4.861059864 & 1.730176500 & $-$1.987324354 \\
$18$      & 0.8143317446 & $-$7.088905249 & 0.9978117866 & $-$6.380628225 & 1.253306995 & $-$5.164338100 & 1.730963611 & $-$2.110590172 \\
$19$      & 0.8132354271 & $-$7.507614324 & 0.9977049784 & $-$6.756988420 & 1.252591524 & $-$5.468002437 & 1.731710905 & $-$2.233774063 \\
$20$      & 0.8138433523 & $-$7.926671808 & 0.9987567634 & $-$7.133067005 & 1.252254212 & $-$5.771915425 & 1.732387532 & $-$2.356901669 \\
\lasthline
\end{tabular}
\begin{tabular}{c|ll|ll|ll|ll}
\firsthline
 &\multicolumn{2}{c|}{$j=0.3$} &\multicolumn{2}{c|}{$j=0.4$}&\multicolumn{2}{c|}{$j=0.5$}&\multicolumn{2}{c}{$j=0.6$} \\
\cline{2-9}
~$n$~    & Re$(\omega_{22n})$ & Im$(\omega_{22n})$ & Re$(\omega_{22n})$ & Im$(\omega_{22n})$ & Re$(\omega_{22n})$ & Im$(\omega_{22n})$ & Re$(\omega_{22n})$ & Im$(\omega_{22n})$ \\
\hline
$0$      & 0.8390533635 & $-$0.1754585438 & 0.8796838435 & $-$0.1737639241 & 0.9282460520 & $-$0.1712776700 & 0.9880895636 & $-$0.1675304043 \\
$1$      & 0.7967806521 & $-$0.5360971407 & 0.8416933545 & $-$0.5294668980 & 0.8948140749 & $-$0.5204491073 & 0.9596133305 & $-$0.5076937291 \\
$2$      & 0.7238544670 & $-$0.9225193799 & 0.7755585095 & $-$0.9065189981 & 0.8358500895 & $-$0.8865737153 & 0.9083581784 & $-$0.8606304030 \\
$3$      & 0.6397722208 & $-$1.338485868 & 0.6977117292 & $-$1.307616228 & 0.7646178009 & $-$1.271072973 & 0.8439532125 & $-$1.226307140\\
$4$      & 0.5601130226 & $-$1.773528714 & 0.6230921623 & $-$1.723476067 & 0.6954825080 & $-$1.665189510 & 0.7798043997 & $-$1.595053108 \\
$5$      & 0.4929144333 & $-$2.213829956 & 0.5633820095 & $-$2.143050811 & 0.6443134678 & $-$2.061346046 & 0.7377914476 & $-$1.962530296 \\
$6$      & 0.44516925 & $-$2.653234973 & 0.5271416128 & $-$2.566946686 & 0.6191370019 & $-$2.467990016 & 0.7242662304 & $-$2.347528328 \\
$7$      & 0.41950589 & $-$3.09783754 & 0.5100960501 & $-$3.003806949 & 0.6087077654 & $-$2.893133945 & 0.7201686625 & $-$2.755612054 \\
$8$      & 0.40782737 & $-$3.55434517 & 0.5011226136 & $-$3.454206260 & 0.6020366776 & $-$3.331983570 & 0.7161323243 & $-$3.177059693 \\
$9$      & 0.40119368 & $-$4.02125475 & 0.4943511668 & $-$3.913403485 & 0.5959004059 & $-$3.778600520 & 0.7112795370 & $-$3.605502978 \\
$10$     & 0.39581575 & $-$4.49463425 & 0.4881585782 & $-$4.377734465 & 0.5898571164 & $-$4.229729924 & 0.7060184617 & $-$4.038079776 \\
$11$     & 0.39070631 & $-$4.97180038 & 0.4822358228 & $-$4.845193461 & 0.5839508719 & $-$4.683743460 & 0.7006677102 & $-$4.473419270 \\
$12$     & 0.38570115 & $-$5.45127817 & 0.4765546912 & $-$5.314713889 & 0.5782475636 & $-$5.139777062 & 0.6953965247 & $-$4.910785820 \\
$13$     & 0.3808149 & $-$5.9322651 & 0.4711244342 & $-$5.785700178 & 0.5727794939 & $-$5.597335053 & 0.6902888849 & $-$5.349747472 \\
$14$     & 0.3760820 & $-$6.4143054 & 0.4659468383 & $-$6.257798707 & 0.5675547305 & $-$6.056113625 & 0.6853895644 & $-$5.790035380 \\
$15$     & 0.3715247 & $-$6.8971262 & 0.461013599 & $-$6.730787346 & 0.5625701103 & $-$6.515917488 & 0.6807305831 & $-$6.231477904 \\
$16$     & 0.367152 & $-$7.380556 & 0.456310557 & $-$7.204519905 & 0.5578201109 & $-$6.976617813 & 0.6763473947 & $-$6.673967229 \\
$17$     & 0.362965 & $-$7.864483 & 0.451821586 & $-$7.678896713 & 0.5533024809 & $-$7.438129856 & 0.6722917794 & $-$7.117441612 \\
$18$     & 0.358955 & $-$8.348830 & 0.447531216 & $-$8.153848286 & 0.5490223226 & $-$7.900400741 & 0.6686471446 & $-$7.561875483 \\
$19$     & 0.355114 & $-$8.833542 & 0.443426261 & $-$8.629325904 & 0.5449961014 & $-$8.363403003 & 0.6655537541 & $-$8.007272028 \\
$20$     & 0.351431 & $-$9.318581 & 0.439496884 & $-$9.105296052 & 0.5412571050 & $-$8.827131870 & 0.6632589339 & $-$8.453648174 \\
\lasthline
\end{tabular}
\end{center}
\caption{QN frequencies $\omega_{22n}$ for $n= 0, 1, 2, ..., 20$ and $j=0.3$, $0.4$, $0.5$, $0.6$, $0.7$, $0.8$, $0.9$, and $0.99$.}
\end{table}

\section{Consistency check of our result for $E_{22n}$}
\label{app:validity}
\begin{figure}[h]
  \begin{center}
    \includegraphics[keepaspectratio=true,height=50mm]{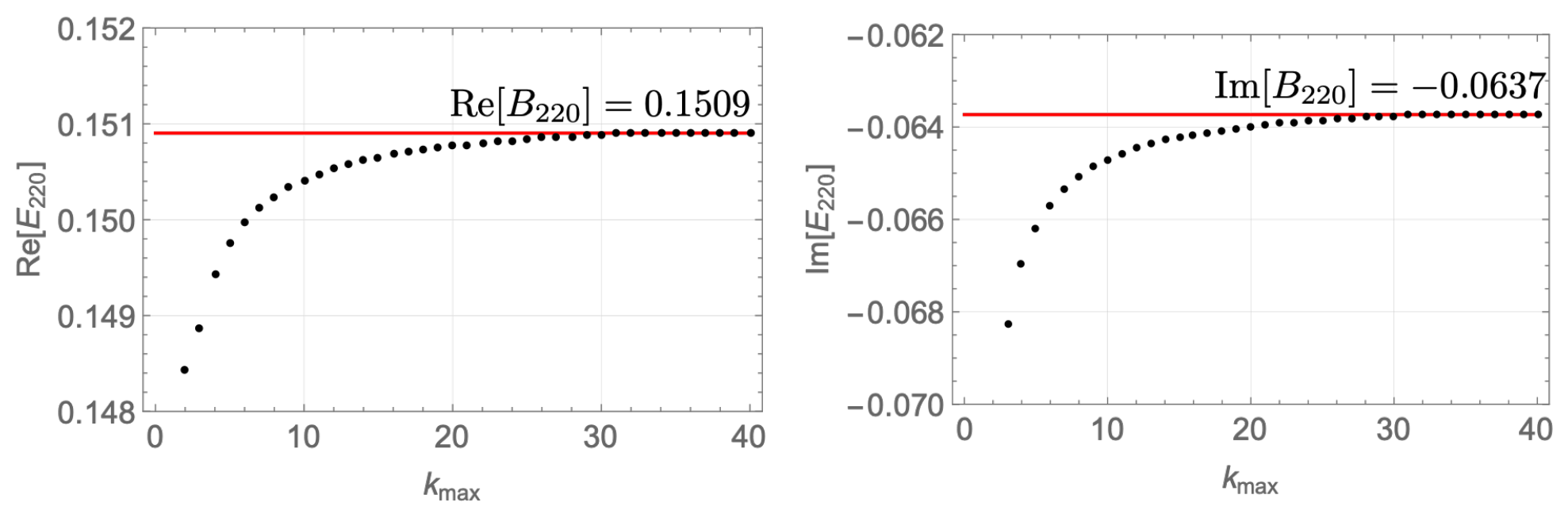}
  \end{center}
\caption{Extrapolated values of $E_{lmn}$ with $(l,m,n) = (2,2,0)$ for $j=0.8$. As $k_{\rm max}$ increases and approaches to $N-1$, the extrapolated value of the QNEF converges to a finite value (indicated by red lines).
}
\label{conv}
\end{figure}

In this Appendix, we show the numerical convergence of the QNEFs $E_{22n}$ in the limit of $\Lambda \to 0$. Also, we check the consistency with the previous study where QNEFs were computed up to the 3rd overtone by Zhang, Berti, and Cardoso \cite{Zhang:2013ksa}. To see the convergence of $E_{22n} (\Lambda)$ extrapolated to the value of $\Lambda = 0$, let us introduce the following fitting function
\begin{equation}
\displaystyle
E_{lmn} (\Lambda) =\sum_{k=0}^{k_{\rm max}} c_k \left( \frac{1}{\ell} \right)^{k}
\label{kmaxex}
\end{equation}
where $k_{\rm max}$ is an integer and $1 \leq k_{\rm max} \leq N-1$. FIG. \ref{conv} shows the values of real and imaginary parts of a QNEF obtained by the extrapolation from (\ref{kmaxex}). One can see that the QNEF converges to a finite value as $k_{\rm max}$ increases. The values of QNEFs obtained in such a way are consistent with the result of the previous research \cite{Zhang:2013ksa} within the error of $\lesssim 0.01 \%$ (see Table \ref{table_zbc}). Note that our definition of the QNEF differs from that in Ref. \cite{Zhang:2013ksa}
\begin{equation}
E^{\rm (ZBC)}_{lmn} = i e^{-i \omega_{lmn}t_s}\omega_{lmn}^2 E_{lmn},
\label{diff_ef}
\end{equation}
where $E^{\rm (ZBC)}_{lmn}$ is the QNEF defined in \cite{Zhang:2013ksa}. In Table \ref{table_zbc}, we show the comparison of $E_{22n}^{\rm (ZBC)}$ in Ref. \cite{Berti:cite,grit:cite} and $E_{22n}^{\rm (ZBC)}$ we computed.
The error of our result, $\Delta_{\rm error}$, is estimated by $\Delta_{\rm error} \equiv \sqrt{\Delta_{\rm abs}^2 + \Delta_{\rm arg}^2}$ where
\begin{equation}
\Delta_{\rm abs} \equiv \frac{\delta |E_{lmn}^{\rm (ZBC)}|}{|E_{lmn}^{\rm (ZBC)}|}, \ 
\Delta_{\rm arg} \equiv \frac{\delta {\rm Arg}(E_{lmn}^{\rm (ZBC)})}{2 \pi}.
\end{equation}
The extra factor of $e^{-i \omega_{lmn} t_s}$ in (\ref{diff_ef}) can be absorbed into the uncertainty of the start time of ringdown (see Eq. (\ref{metric_perturb_ET})). Therefore, we should look into the consistency between the QNEFs obtained in \cite{Zhang:2013ksa} and ours by taking into account the ambiguity of $e^{-i \omega_{lmn} t_s}$.
%%%%%%%%%%%%%%%%%%%%%%%%%
%%%%%%%%%%%%%%%%%%%%%%%%%

\begin{table}
\begin{center}
\begin{tabular}{c|cc|cc|c|c}
\firsthline
spin and overtone number &\multicolumn{2}{c|}{our work} &\multicolumn{2}{c|}{Z. Zhang et al. \cite{Zhang:2013ksa,Berti:cite,grit:cite}}& error & start time \\
\cline{2-7}
$(a,n)$    & $|E_{lmn}^{\rm (ZBC)}|$ & arg$(E_{lmn}^{\rm (ZBC)})$ & $|E_{lmn}^{\rm (ZBC)}|$ & arg$(E_{lmn}^{\rm (ZBC)})$ & $\Delta_{\rm error} [\%]$ & $t_s$ \\
\hline
$j=0.3, n=0$      & 0.056048    &  $-$0.07956  & 0.056045 & $-$0.07940& 0.007 & 0.146100  \\
$j=0.3, n=1$      & 0.22975        & 3.070    & 0.22972& 3.070 & 0.01 & 0.146100 \\
$j=0.3, n=2$      & 0.65693     & $-$0.09014   & 0.65689& $-$0.09003 & 0.006 & 0.146100 \\
$j=0.3, n=3$      & 1.5678     & 3.043   & 1.5679& 3.043 & 0.005 & 0.146100 \\
\hline
$j=0.4, n=0$      & 0.071607    & 0.1813   & 0.071600 & 0.1812 & 0.01 & 0.262800 \\
$j=0.4, n=1$      & 0.31881        & $-$2.895    & 0.31877& $-$2.895 & 0.01 & 0.262800\\
$j=0.4, n=2$      & 0.97028     & 0.2671   & 0.97019& 0.2671 & 0.01  & 0.262800\\
$j=0.4, n=3$      & 2.4422     & $-$2.864   & 2.4421& $-$2.864  & 0.005 & 0.262800\\
\hline
$j=0.5, n=0$      & 0.094264    & 0.5181   & 0.094260 & 0.5181  & 0.004 & 0.424605\\
$j=0.5, n=1$      & 0.46338        & $-$2.487    & 0.46336& $-$2.487 & 0.004 & 0.424605\\
$j=0.5, n=2$      & 1.5233     & 0.7295   & 1.5232& 0.7296 & 0.003 & 0.424605\\
$j=0.5, n=3$      & 4.0992     & $-$2.369   & 4.0992& $-$2.369 & 0.0005 & 0.424605 \\
\hline
$j=0.6, n=0$      & 0.128880    & 0.9684  & 0.128875 & 0.9684 & 0.004 &0.649046 \\
$j=0.6, n=1$      & 0.71357    & $-$1.945    & 0.71354& $-$1.945 & 0.004 &0.649046 \\
$j=0.6, n=2$      & 2.5803     & 1.350   & 2.5802& 1.350 & 0.003 &0.649046 \\
$j=0.6, n=3$      & 7.5714     & $-$1.6954   & 7.5714& $-$1.6953 & 0.001 &0.649046 \\
\hline
$j=0.7, n=0$      & 0.18508    & 1.604   & 0.18507 & 1.603 & 0.01 & 0.962042 \\
$j=0.7, n=1$      & 1.1820        & $-$1.185    & 1.1819& $-$1.185 & 0.005 & 0.962042\\
$j=0.7, n=2$      & 4.8059     & 2.228   & 4.8058& 2.228 & 0.003 & 0.962042\\
$j=0.7, n=3$      & 15.859     & $-$0.7214   & 15.859& $-$0.7213 & 0.001 & 0.962042\\
\hline
$j=0.8, n=0$      & 0.28378    & 2.584   & 0.28376 & 2.584 & 0.007 & 1.424616\\
$j=0.8, n=1$      & 2.1453        & $-$0.02295 & 2.1452 & $-$0.02289  & 0.006 & 1.424616\\
$j=0.8, n=2$      & 9.9927     & $-$2.696   & 9.9924& $-$2.696  & 0.004 & 1.424616\\
$j=0.8, n=3$      & 38.437 & 0.8425 & 38.437 & 0.8427 & 0.002 & 1.424616 \\
\hline
$j=0.9, n=0$      & 0.47310    & $-$1.856   & 0.47307 & $-$1.856  & 0.008 &2.227500 \\
$j=0.9, n=1$      & 4.2845        & 2.113    & 4.2842& 2.113 & 0.006 &2.227500\\
$j=0.9, n=2$      & 22.362     & $-$0.2090   & 22.362& $-$0.2088 & 0.004 &2.227500\\
$j=0.9, n=3$      & 92.437     & $-$2.5103   & 92.437& $-$2.5101 & 0.003  &2.227500\\
\hline
$j=0.99, n=0$      & 0.59492    & $-$1.596   & 0.59483 & $-$1.595 & 0.02 & 4.778240 \\
$j=0.99, n=1$      & 5.0958        & 2.992    & 5.0947& 2.992 & 0.02 & 4.778240\\
$j=0.99, n=2$      & 23.029     & 1.412   & 23.028 & 1.413 & 0.01 & 4.778240\\
$j=0.99, n=3$      & 74.329     & $-$0.06519   & 74.329& $-$0.06449 & 0.01 & 4.778240 \\
\lasthline
\end{tabular}
\end{center}
\caption{Comparison with the excitation factors computed by Zhang, Berti, and Cardoso in Ref. \cite{Zhang:2013ksa}.}
\label{table_zbc}
\end{table}

\section{Source term of a particle plunging into a black hole}
\label{app:source}

In Sec. \ref{sec:fit_overtones}, we compute GW signals induced by a particle plunging into a spinning black hole. The source term in terms of the Sasaki-Nakamura formalism is given by \cite{Kojima:1984cj,Nakamura:1987zz}
\begin{equation}
\tilde{T}^{\rm (SN)}_{lm} = \frac{\gamma \Delta \mu \tilde{W}}{(r^2+a^2)^{3/2} r^2} \exp \left( -i k r^{\ast} \right),
\end{equation}
with
\begin{align}
\tilde{W} &\equiv W_{nn} + W_{n\bar{m}} + W_{\bar{m} \bar{m}},\\
W_{nn} &\equiv f_0 e^{i \chi} + \int^{\infty}_{r} dr' f_1 e^{i \chi} + \int^{\infty}_r dr' \int^{\infty}_{r'} dr'' f_2 e^{i\chi},\\
W_{n \bar{m}} &\equiv g_0 e^{i \chi} + \int^{\infty}_{r} dr' g_1 e^{i\chi},\\
W_{\bar{m} \bar{m}} &\equiv h_0 e^{i\chi} + \int^{\infty}_r dr' h_1 e^{i \chi} + \int^{\infty}_r dr' \int^{\infty}_{r'} dr'' h_2 e^{i\chi},
\end{align}
where $\chi \equiv \omega t - m\varphi + k r^{\ast}$ and
\begin{align}
f_0 &\equiv - \frac{1}{\omega} \frac{r^2 \sqrt{\cal R}}{(r^2+a^2)^2} S_2,\\
f_1 &\equiv \frac{f_0}{S_c} \left[ (S_1 + (a \omega-m) S_0) \frac{ia}{r^2} + S_2 \left\{ \frac{2 (a^2-r^2)}{r (r^2+a^2)} + \frac{{\cal R}'}{2 {\cal R}} + i \eta \right\} \right],\\
f_2 &\equiv \frac{i}{\omega} \frac{r^2 \sqrt{\cal R}}{(r^2+a^2) \Delta} \left( 1 - \frac{P}{\sqrt{\cal R}} \right) \left[ \left\{ S_1 + (a \omega -m) S_0 \right\} \frac{ia}{r^2} +S_2 \left\{ \frac{2a^2}{r (r^2+a^2)} + \frac{2r}{r^2+(L-a)^2} -\frac{(P+\sqrt{R})'}{P + \sqrt{R}} + i\eta \right\} \right],\\
\eta &\equiv \frac{(a\omega -m)(a-L)}{\sqrt{\cal R}} - \frac{am}{\Delta} \left( 1- \frac{P}{\sqrt{\cal R}} \right),\\
g_0 &\equiv - \frac{a-L}{\omega} \left\{ S_1 + (a\omega - m) S_0 \right\} \frac{r^2}{r^2 + a^2},\\
g_1 &\equiv g_0 \left[ \frac{2a^2}{r (r^2+a^2)} + i \eta \right],\\
h_0 &\equiv - \frac{r^2 h_2}{2},\\
h_1 &\equiv -rh_2,\\
h_2 &\equiv \frac{S_0 (a-L)^2}{\sqrt{\cal R}},\\
S_0 &\equiv {}_{-2} S_{lm} (\pi/2),\\
S_1 &\equiv \left. \frac{d}{d \theta} {}_{-2} S_{lm} (\theta) \right|_{\theta = \pi/2},\\
S_2 &\equiv \left( a \omega - m - \frac{ia}{r} \right) \left[ S_1 + (a \omega-m) S_0 \right] - \frac{\lambda}{2} S_0.
\end{align}

\section{Details of the fitting analysis in Sec. \ref{sec:fit_overtones}}
\label{app:fit}
\begin{figure}[H]
  \begin{center}
    \includegraphics[keepaspectratio=true,height=40mm]{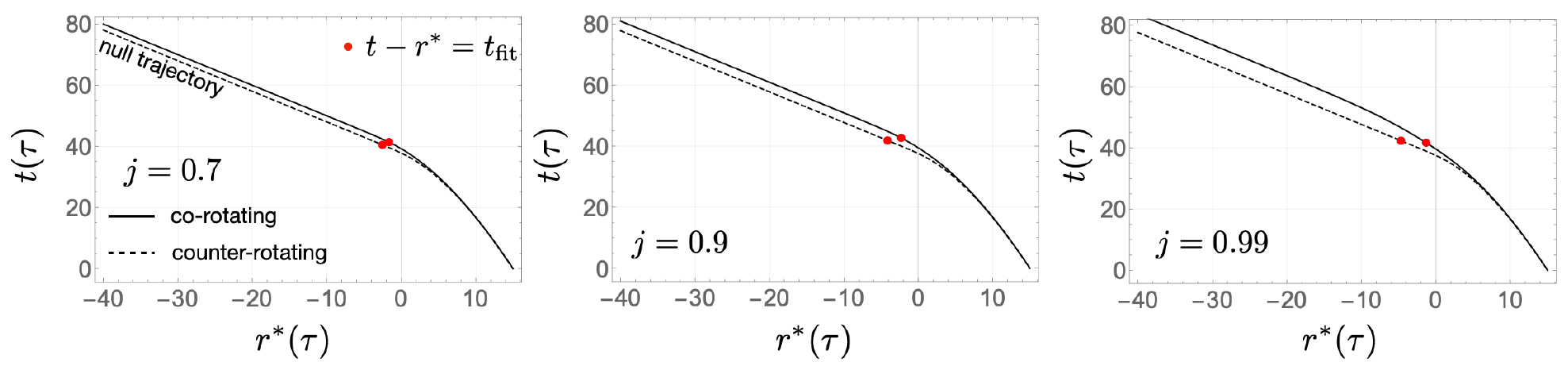}
  \end{center}
\caption{Trajectory of the particle plunging into a black hole ($r^{\ast} (\tau), t(\tau)$). Red points indicate the moment when the particle approaches to the horizon and starts to follow the null geodesics.
}
\label{t_rast}
\end{figure}

In Sec. \ref{sec:fit_overtones}, we fit the QN modes up to the 7th or 11th overtone to the numerical GW waveform induced by a particle plunging into a spinning black hole. We set the fit start time to the moment when the particle approaches to the horizon and starts to follow the null geodesics. In this case, the trajectory of the particle is almost null in the tortoise coordinate (FIG. \ref{t_rast}), which may be regarded as the absorption of the particle by the black hole.

\begin{figure}[H]
  \begin{center}
    \includegraphics[keepaspectratio=true,height=50mm]{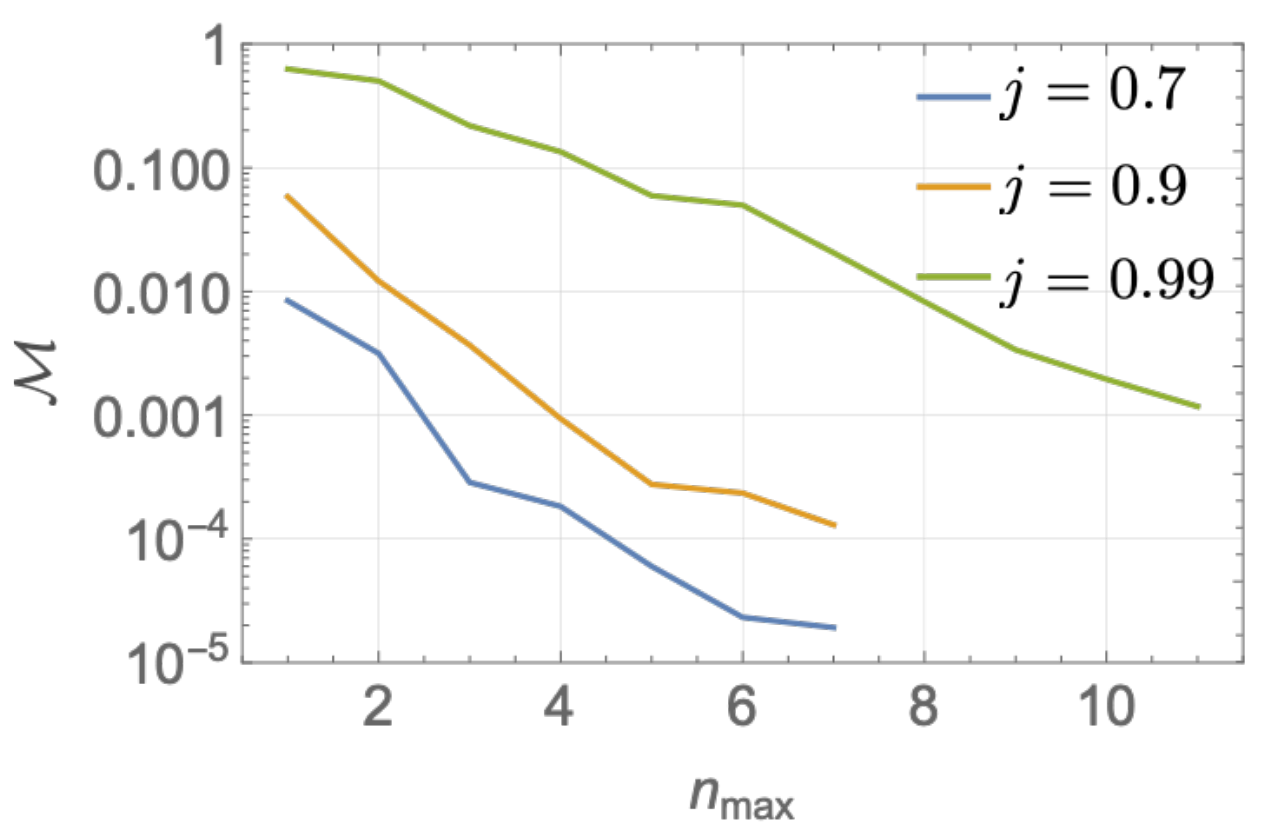}
  \end{center}
\caption{Mismatches for $j=0.7$, $0.9$, and $0.99$ with $L=1$.
}
\label{mismatch}
\end{figure}

To see the significance of the inclusion of overtones, we compute the mismatch, ${\cal M}$, between the numerical GW waveforms, $h$, and the waveforms modeled by the superposition of QN modes, $h_{Q}$, while changing the number of overtones included in $h_{Q}$ denoted by $n_{\rm max}$ (FIG. \ref{mismatch}), where ${\cal M}$ is defined as
\begin{equation}
{\cal M} \equiv \left| 1- \frac{\braket{h|h_Q}}{\sqrt{\braket{h|h} \braket{h_Q | h_Q}}} \right|,
\end{equation}
with
\begin{equation}
\braket{A|B} \equiv \int dt A(t) B^{\ast}(t).
\end{equation}
Taking many overtones into account is necessary to reduce the mismatch, ${\cal M}$, for higher spin parameters (see FIG. \ref{ampls}), it may cause a hierarchy of ${\cal M}(j=0.7) < {\cal M}(j= 0.9) < {\cal M} (j=0.99)$ for a fixed $n_{\rm max}$ as shown in FIG. \ref{mismatch}.
%%%%%%%%%%%%%%%%%%%%%%%%%
%%%%%%%%%%%%%%%%%%%%%%%%%
%%%%%%%%%%%%%%%%%%%%%%%%%
%%%%%%%%%%%%%%%%%%%%%%%%%

\end{document}